\documentclass[5p,sort&compress]{elsarticle}
\usepackage{amssymb}
\usepackage{amsmath}
\usepackage{booktabs}
\usepackage{multirow}
\usepackage{calc}
\usepackage{threeparttable}
\usepackage[hidelinks]{hyperref}

\journal{Astroparticle Physics}

\newcommand{\gls}[1]{#1}

\newcommand{\cena}{Centaurus~A}

\newcommand{\specialcell}[2]{ \begin{tabular}{@{}#1@{}}#2\end{tabular} }
\renewcommand{\tnote}[1]{$^{\textrm #1}$}
\newcommand{\titem}[1]{\item \tnote{#1} \footnotesize}
\newlength{\tnotealen}
\newlength{\tnoteblen}
\newlength{\tnoteclen}
\newlength{\tnotedlen}
\newcommand{\sinc}{{\rm sinc}}

\newcommand{\degree}{\ensuremath{^{\circ}}}
\newcommand{\arcmin}{\ensuremath{^{\prime}}}

\newcommand{\eqn}{Eq.}
\newcommand{\eqns}{Eqs.}
\newcommand{\fig}{Fig.}

\newcommand{\tab}{Table}

\newcommand{\sect}{Sec.}
\newcommand{\sects}{Secs.}

\newcommand{\Fig}{Fig.}

\newcommand{\Sect}{Sec.}

\newcommand{\refii}[2]{\ref{#1} and \ref{#2}}
\newcommand{\refiii}[3]{\ref{#1}, \ref{#2} and \ref{#3}}
\newcommand{\refs}[2]{\ref{#1}--\ref{#2}}

\newcommand{\eqnref}[1]{\eqn{}~\ref{#1}}

\newcommand{\eqnrefii}[2]{\eqns{}\ \refii{#1}{#2}}

\newcommand{\figref}[1]{\fig{}~\ref{#1}}
\newcommand{\Figref}[1]{\Fig{}~\ref{#1}}

\newcommand{\tabref}[1]{\tab{}~\ref{#1}}

\newcommand{\secref}[1]{\sect{}~\ref{#1}}
\newcommand{\Secref}[1]{\Sect{}~\ref{#1}}
\newcommand{\secrefii}[2]{\sects{}\ \refii{#1}{#2}}

\newcommand{\secrefiii}[3]{\sects{}\ \refiii{#1}{#2}{#3}}

\newcommand{\secrefs}[2]{\sects{}\ \refs{#1}{#2}}

\newcommand{\gcitep}{~\citep}

\newcounter{cutenumi}
\newenvironment{cutenumerate}{ \begin{enumerate}[\#1:] \setcounter{enumi}{\value{cutenumi}} }{ \setcounter{cutenumi}{\value{enumi}} \end{enumerate} }
\newcommand{\cutref}[1]{cut~\#\ref{#1}}
\newcommand{\Cutref}[1]{Cut~\#\ref{#1}}
\newcommand{\cutsref}[2]{cuts \mbox{\#\ref{#1}--\ref{#2}}}
\newcommand{\Cutsref}[2]{Cuts \mbox{\#\ref{#1}--\ref{#2}}}

\begin{document}

\newcommand{\aap}{A\&A}                  
\newcommand{\aapr}{A\&A Rev.}            
\newcommand{\aaps}{A\&AS}                
\newcommand{\aipcs}{AIP Conf.\ Series}   
\newcommand{\aj}{AJ}                     
\newcommand{\ajph}{Australian J.\ Phys.} 
\newcommand{\alet}{Astro.\ Lett.}        
\newcommand{\anchem}{Analytical Chem.}   
\newcommand{\ao}{Applied Optics}         
\newcommand{\apj}{ApJ}                   
\newcommand{\apjl}{ApJ Lett.}                  
\newcommand{\apjs}{ApJS}                 
\newcommand{\app}{Astropart.\ Phys.}     
\newcommand{\apss}{Ap\&SS}               
\newcommand{\apssproc}{Ap\&SS\ Proc.}    
\newcommand{\araa}{ARA\&A}               
\newcommand{\arep}{Astron.\ Rep.}        
\newcommand{\arxiv}{ArXiv e-prints}      
\newcommand{\aspacer}{Adv.\ Space Res.}  
\newcommand{\aspconf}{Astron.\ Soc.\ Pac.\ Conf.} 
\newcommand{\atel}{ATel}                 
\newcommand{\azh}{AZh}                   
\newcommand{\baas}{BAAS}                 
\newcommand{\bell}{Bell Systems Tech.\ J.} 
\newcommand{\cpc}{Comput.\ Phys.\ Commun.} 
\newcommand{\cosres}{Cosm.\ Res.}        
\newcommand{\dans}{Dokl.\ Akad.\ Nauk SSSR}  
\newcommand{\elec}{Electronics}          
\newcommand{\epjwoc}{EPJ Web of Conf.}   
\newcommand{\epsl}{Earth and Plan.\ Sci.\ Lett.} 
\newcommand{\expa}{Exp.\ Astron.}        
\newcommand{\gca}{Geochim.\ Cosmochim.\ Acta} 
\newcommand{\grl}{Geophys.\ Res.\ Lett.} 
\newcommand{\iaucirc}{IAU Circ.}         
\newcommand{\iauproc}{Proc.\ of the IAU} 
\newcommand{\ibvs}{IBVS}                 
\newcommand{\icarus}{Icarus}             
\newcommand{\ieeetit}{IEEE Trans.\ Inf.\ Theor.} 
\newcommand{\ieeemtt}{IEEE Trans.\ Microwave Theor.\ \& Techniques} 
\newcommand{\ijmpd}{Int'l J.\ Mod.\ Phys.\ D} 
\newcommand{\invp}{Inverse Prob.}        
\newcommand{\jastp}{J.\ Atmos.\ Sol.-Terr.\ Phys.} 
\newcommand{\jcap}{J.\ Cosm.\ Astropart.\ Phys.} 
\newcommand{\jcomph}{J.\ Comput.\ Phys.} 
\newcommand{\jcp}{J.\ Chem.\ Phys.}      
\newcommand{\jewa}{J.\ Electromagn.\ Wav.\ Appl} 
\newcommand{\jgeod}{J.\ Geodesy}         
\newcommand{\jgr}{J.\ Geophys.\ Res.}    
\newcommand{\jhep}{JHEP}                 
\newcommand{\jrasc}{JRASC}               
\newcommand{\met}{Meteoritics}           
\newcommand{\mmras}{MmRAS}               
\newcommand{\mnras}{MNRAS}               
\newcommand{\moonp}{Moon and Plan.}      
\newcommand{\mpla}{Mod.\ Phys.\ Lett.\ A} 
\newcommand{\mps}{Meteoritics and Planetary Science} 
\newcommand{\nar}{New Astron.\ Rev.}     
\newcommand{\nast}{New Astron.}          
\newcommand{\nat}{Nature}                
\newcommand{\nima}{Nucl.\ Instrum.\ Meth.\ A} 
\newcommand{\npbproc}{Nucl.\ Phys.\ B Proc.\ Supp.} 
\newcommand{\njp}{New J.\ Phys.}         
\newcommand{\nspu}{Phys.\ Uspekhi}       
\newcommand{\pasa}{PASA}                 
\newcommand{\pasj}{PASJ}                 
\newcommand{\pasp}{PASP}                 
\newcommand{\phr}{Phys.\ Rev.}           
\newcommand{\pla}{Phys.\ Lett.\ A}       
\newcommand{\plb}{Phys.\ Lett.\ B}       
\newcommand{\pop}{Phys.\ Plasmas}        
\newcommand{\pra}{Phys.\ Rev.\ A}        
\newcommand{\prb}{Phys.\ Rev.\ B}        
\newcommand{\prc}{Phys.\ Rev.\ C}        
\newcommand{\prd}{Phys.\ Rev.\ D}        
\newcommand{\pre}{Phys.\ Rev.\ E}        
\newcommand{\prl}{Phys.\ Rev.\ Lett.}    
\newcommand{\pst}{Phys.\ Scr.\ T}        
\newcommand{\phrep}{Phys.\ Rep.}         
\newcommand{\phss}{Phys.\ Stat.\ Sol.}   %
\newcommand{\privcom}{priv.\ comm.}      
\newcommand{\procsci}{Proc.\ Sci.}       
\newcommand{\procspie}{Proc.\ SPIE}      
\newcommand{\planss}{Planet.\ Space Sci.} 
\newcommand{\qjras}{QJRAS}               
\newcommand{\radsci}{Radio Sci.}         
\newcommand{\rpph}{Rep.\ Prog.\ Phys.}   
\newcommand{\rqe}{Rad.\ \& Quan.\ Elec.} 
\newcommand{\rgsp}{Rev.\ Geophys.\ Space Phys.\ } 
\newcommand{\rsla}{Phil.\ Trans.\ R.\ Soc.\ A} 
\newcommand{\sal}{Sov.\ Astron.\ Lett.}  
\newcommand{\spjetp}{Sov.\ Phys.\ JETP}  
\newcommand{\spjetpl}{Sov.\ Phys.\ JETP Lett.} 
\newcommand{\spu}{Sov.\ Phys.\ Usp.}  
\newcommand{\sci}{Science}               
\newcommand{\solph}{Sol.\ Phys.}         
\newcommand{\ssr}{Space Sci.\ Rev.}      
\newcommand{\zap}{Z.\ Astrophys.}        

\begin{frontmatter}

\title{A lunar radio experiment with the Parkes radio telescope for the LUNASKA project}

\author[adelaideuni,atnf,sotonuni]{J.D.\ Bray\corref{cor1}}
\cortext[cor1]{Corresponding author.}
\ead{j.bray@soton.ac.uk}

\author[atnf]{R.D.\ Ekers}
\author[atnf]{P.\ Roberts}
\author[atnf]{J.E.\ Reynolds}
\author[erlangenuni]{C.W.\ James}
\author[atnf]{C.J.\ Phillips}
\author[adelaideuni]{R.J.\ Protheroe}
\author[astron]{R.A.\ McFadden}
\author[adelaideuni]{M.G.\ Aartsen}

\address[adelaideuni]{School of Chemistry \& Physics, Univ.\ of Adelaide, SA 5005, Australia}
\address[atnf]{CSIRO Astronomy \& Space Science, Epping, NSW 1710, Australia}
\address[sotonuni]{School of Physics \& Astronomy, Univ.\ of Southampton, SO17 1BJ, United Kingdom}
\address[erlangenuni]{ECAP, Univ.\ of Erlangen-Nuremberg, 91058 Erlangen, Germany}
\address[astron]{ASTRON, 7990 AA Dwingeloo, The Netherlands}

\begin{abstract}
 We describe an experiment using the Parkes radio telescope in the 1.2--1.5~GHz frequency range as part of the LUNASKA project, to search for nanosecond-scale pulses from particle cascades in the Moon, which may be triggered by ultra-high-energy astroparticles.  Through the combination of a highly sensitive multi-beam radio receiver, a purpose-built backend and sophisticated signal-processing techniques, we achieve sensitivity to radio pulses with a threshold electric field strength of 0.0053 $\mu$V/m/MHz, lower than previous experiments by a factor of three.  We observe no pulses in excess of this threshold in observations with an effective duration of 127 hours.  The techniques we employ, including compensating for the phase, dispersion and spectrum of the expected pulse, are relevant for future lunar radio experiments.
\end{abstract}


\end{frontmatter}

\section{Introduction}

Studies of ultra-high-energy (UHE; \mbox{$> 10^{18}$}~eV) cosmic rays, and searches for their expected counterpart neutrinos, are difficult because of their extremely low flux.  Current observatories such as the Pierre Auger Observatory\gcitep{abraham2004} and the Telescope Array\gcitep{abu-zayyad2012c} make use of arrays of detectors covering thousands of km$^2$ to detect the particle cascades produced when UHE particles interact in the Earth's atmosphere.  An alternative approach\gcitep{dagkesamanskii1989} is to use the Moon as the detector, searching for the nanosecond-scale radio pulse produced when a UHE particle interacts in a dense medium such as the lunar regolith\gcitep{askaryan1962} with terrestrial radio telescopes.  This was originally proposed as a means to detect UHE neutrinos, but is also sensitive to cosmic rays\gcitep{james2009b,terveen2010,jeong2012}, and the large size of the Moon gives it a larger potential aperture than other detection techniques.  The detection of an individual nanosecond-scale pulse requires techniques that are quite unlike those of conventional radio astronomy, and are being refined through ongoing experimentation, with the ultimate goal of establishing lunar radio observations as a practical technique for detecting UHE particles.

The first lunar particle detection experiment was performed in 1995 with the Parkes radio telescope\gcitep{hankins1996}; subsequent experiments have been carried out with the Goldstone Deep Space Communications Complex (GDSCC) \citep[GLUE]{gorham2004a}, the Kalyazin radio telescope\gcitep{beresnyak2005}, the Australia Telescope Compact Array (ATCA) \citep[LUNASKA]{james2010}, the Lovell telescope \citep[La Luna]{spencer2010}, the Westerbork Synthesis Radio Telescope (WSRT) \citep[NuMoon]{buitink2010} and the Expanded Very Large Array (EVLA) \citep[RESUN]{jaeger2010}.  A key feature of these experiments is that they make use of existing radio telescopes, whereas other experiments searching for UHE cosmic rays or neutrinos require the development of expensive dedicated instruments.  The sensitivity of radio telescopes is undergoing rapid improvement, through the upgrading of existing instruments and the construction of new ones, which can be exploited to achieve greater sensitivity to UHE particles.  These efforts are expected to culminate in the Square Kilometre Array (SKA)\gcitep{carilli2004}, an international radio telescope scheduled to begin construction in 2016, which will be sensitive even to the minimal neutrino flux expected from the Greisen-Zatsepin-Kuzmin\gcitep{greisen1966,zatsepin1966} interactions of propagating cosmic rays\gcitep{james2009b}.  The LUNASKA (Lunar Ultra-high energy Neutrino Astrophysics with the SKA) project aims to develop the lunar radio approach to particle detection for future use with the SKA.

Our previous LUNASKA experiment searched for coincident pulses independently detected by three antennas of the ATCA, and was therefore limited by the sensitivity of a single 22~m antenna.  Improving on this with the same instrument would require combining the signals from multiple antennas into tied-array beams, similar to the NuMoon experiment with the WSRT, which would be prohibitively difficult given the ATCA's higher bandwidth and angular resolution.  Instead, we conducted a further experiment with the 64~m Parkes radio telescope, with its greater collecting area allowing an improvement in sensitivity.  The disadvantage of this larger antenna is that its beam observes a smaller fraction of the Moon, which we counteracted by using a multibeam receiver to observe multiple regions of the Moon simultaneously as suggested by \citet{ekers2009}.  The use of a single antenna also required a different strategy from the ATCA experiment for excluding pulses from radio frequency interference (RFI), which were identified primarily by their simultaneous appearance in multiple beams.

In this article, we describe our lunar radio experiment with the Parkes radio telescope in 2010 as part of the LUNASKA project, with improved sensitivity over our previous experiment with the ATCA.  \Secref{sec:design} contains an overview of the design of the experiment.  In \secrefiii{sec:calibration}{sec:ionosphere}{sec:signal} we go into more detail respectively about the calibration procedure, the effects of ionospheric dispersion, and the optimisation of the signal-to-noise ratio.  In \secref{sec:obs} we describe the analysis of the data from our observations, including the identification and exclusion of RFI, and in \secref{sec:conclusion} we conclude by summarising our results and considerations for future experiments.  The limits that our experiment establishes on the ultra-high-energy particle flux will be addressed in a separate work\gcitep{parkes_theory}.

\section{Experiment design}
\label{sec:design}

The Parkes radio telescope (33\degree\,00\arcmin\,S 148\degree\,16\arcmin\,E) is located in New South Wales, Australia, and consists of a single 64~m parabolic antenna with receivers mounted at the prime focus.  This experiment used the 21~cm multibeam receiver\gcitep{staveley-smith1996}, which is capable of observing up to thirteen points on the sky simultaneously.  At any one time we used four of the receiver's thirteen beams, each with two orthogonal linear polarisations; see \secref{sec:pointing} for details of our pointing strategy.  The receiver has a radio frequency (RF) band of 1.2--1.5~GHz, which is downconverted to an intermediate frequency (IF) band of 50--350~MHz.

\begin{figure}
 \centering
 \includegraphics[width=\linewidth]{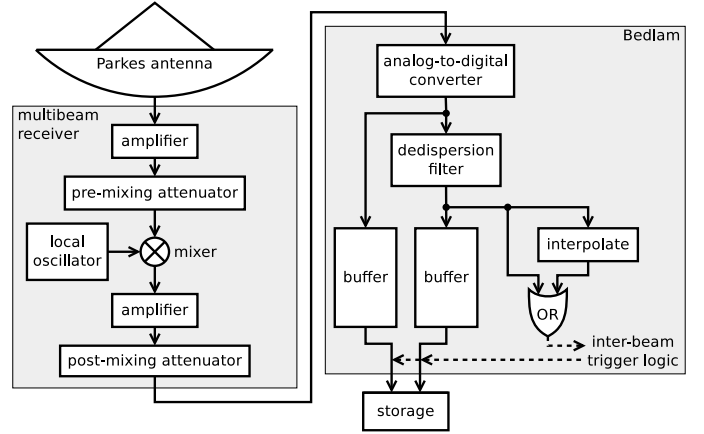}
 \caption{Signal path for a single telescope beam for this experiment, for one polarisation only.  Left side is the signal path through the Parkes 21~cm multibeam receiver, which is standard except for the addition of an attenuator prior to the mixer (see \secref{sec:linearity}).  Right side shows the Bedlam backend built for this experiment.  Anticoincidence trigger logic between the beams is shown in \figref{fig:trigger_logic}.}
 \label{fig:sigpath}
\end{figure}

Further processing of the signal was performed with Bedlam, a digital backend developed specifically for this experiment\gcitep{bray2012}.  The voltage waveform was digitised with 8~bits of precision at a rate of 1,024 Msamples/s, oversampling the band by a factor \mbox{$\sim 1.7$}.  A copy of this digital signal was then passed through a digital filter to compensate for ionospheric dispersion (see \secrefii{sec:ionosphere}{sec:dedispersion}); both the raw and dedispersed data were stored in temporary buffers, with an adjustable size of up to 8~$\mu$s.  \Figref{fig:sigpath} shows details of the signal path.

\begin{figure}
 \centering
 \includegraphics[width=0.65\linewidth]{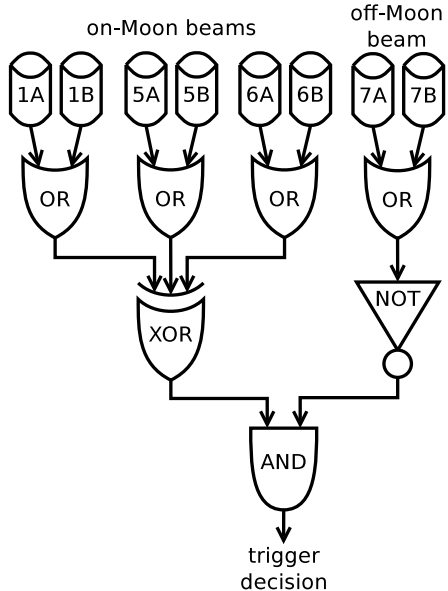}
 \caption{Real-time inter-beam trigger logic, for polarisations A and B of each of four beams.  The choice of beams here is for pointing configuration~A (see \figref{fig:configs}).  In some configurations, two of the four beams were placed off-Moon, but the real-time trigger logic was not changed to reflect this.}
 \label{fig:trigger_logic}
\end{figure}

Due to the high data rate, the buffered data were copied to permanent storage only if triggered by a set of conditions corresponding to the possible detection of a lunar Askaryan pulse.  For a valid trigger, we required an excursion beyond an adjustable threshold in the dedispersed data for a beam pointing at the Moon, in either polarisation; this excursion could occur either in the direct voltage samples, or in an equal number of values interpolated midway between them.  Further, we required that the pulse appear in only a single beam: simultaneous excursions in multiple beams are characteristic of transient RFI from artificial sources in the vicinity of the telescope, detected through the far sidelobes of the beam patterns.  The trigger logic of this anticoincidence filter is illustrated in \figref{fig:trigger_logic}.

The raw trigger rate was highly variable and typically dominated by RFI, with a mean rate of over 200~Hz.  The anticoincidence filter rejected the majority of these raw triggers, with the accepted trigger rate typically in the range 1--2~Hz, dominated by thermal noise.  On an accepted trigger, the raw and dedispersed buffers for all beams were copied to permanent storage; these constitute a single frame of data.  During this data transfer, the system was unable to respond to further triggers, resulting in a minor loss of effective observing time.  The recorded data were subjected to retrospective processing to optimise the signal-to-noise ratio (\secref{sec:signal}), to remove minor instrumental effects (\secref{sec:processing}), and to exclude RFI pulses which escaped the real-time anticoincidence filter (\secref{sec:cuts}).

\subsection{Pointing strategy}
\label{sec:pointing}

The Parkes 21~cm multibeam receiver has a feed consisting of thirteen horn antennas in a hexagonal array, forming a corresponding pattern of beams on the sky.  The feed array can be rotated within a limited range, which is typically used to maintain a constant parallactic angle but can be arbitrarily controlled with software developed for this experiment.  Approximating the telescope antenna as a uniformly-illuminated 64~m aperture (see \secref{sec:calibration}), each beam is a frequency-dependent Airy disk with a full-width half-maximum (FWHM) size of 12.3\arcmin\ at the centre of the band.

Our pointing strategy was dictated by limitations on the geometry of a detectable UHE particle interaction.  The radiation from a particle cascade is directed forward as a hollow cone at the Cherenkov angle; combined with internal reflection at the regolith-vacuum interface, this prevents the radiation from escaping for particles which are steeply down-going with respect to the lunar surface.  Particles, even neutrinos, which are steeply up-going are strongly attenuated as they pass through the Moon.  The only detectable particles are those which intersect the lunar surface at a shallow angle, with the refracted radio emission directed slightly above the surface.  As viewed from the Earth, the Askaryan pulse comes from the Moon's apparent edge or limb, with the initial particle originating from a point typically 15--30\degree\ in the corresponding direction from the Moon\gcitep{james2009f}.  The component of the pulse which escapes the lunar surface has linear polarisation which is typically radial with respect to the Moon.

Combined with the consideration that the noise in our receiver is dominated by thermal radiation from the Moon, these limitations lead us to the following requirements.
\begin{enumerate}
 \item As many beams as possible should be directed towards the limb of the Moon. \label{it:pointreq_limb}
 \item The Moon should occupy as little of each beam as possible, to minimise the thermal lunar noise in the receiver. \label{it:pointreq_moon}
 \item One linear polarisation of each beam should be oriented radially to the Moon, to match the polarisation of the expected signal. \label{it:pointreq_pol}
 \item To maximise our sensitivity to a potential UHE particle source, it should be close (\mbox{$\lesssim 45$}\degree) to the Moon, and one beam should be directed at the point on the lunar limb closest to it. \label{it:pointreq_cena}
\end{enumerate}

\begin{figure}
 \centering
 \includegraphics[width=\linewidth]{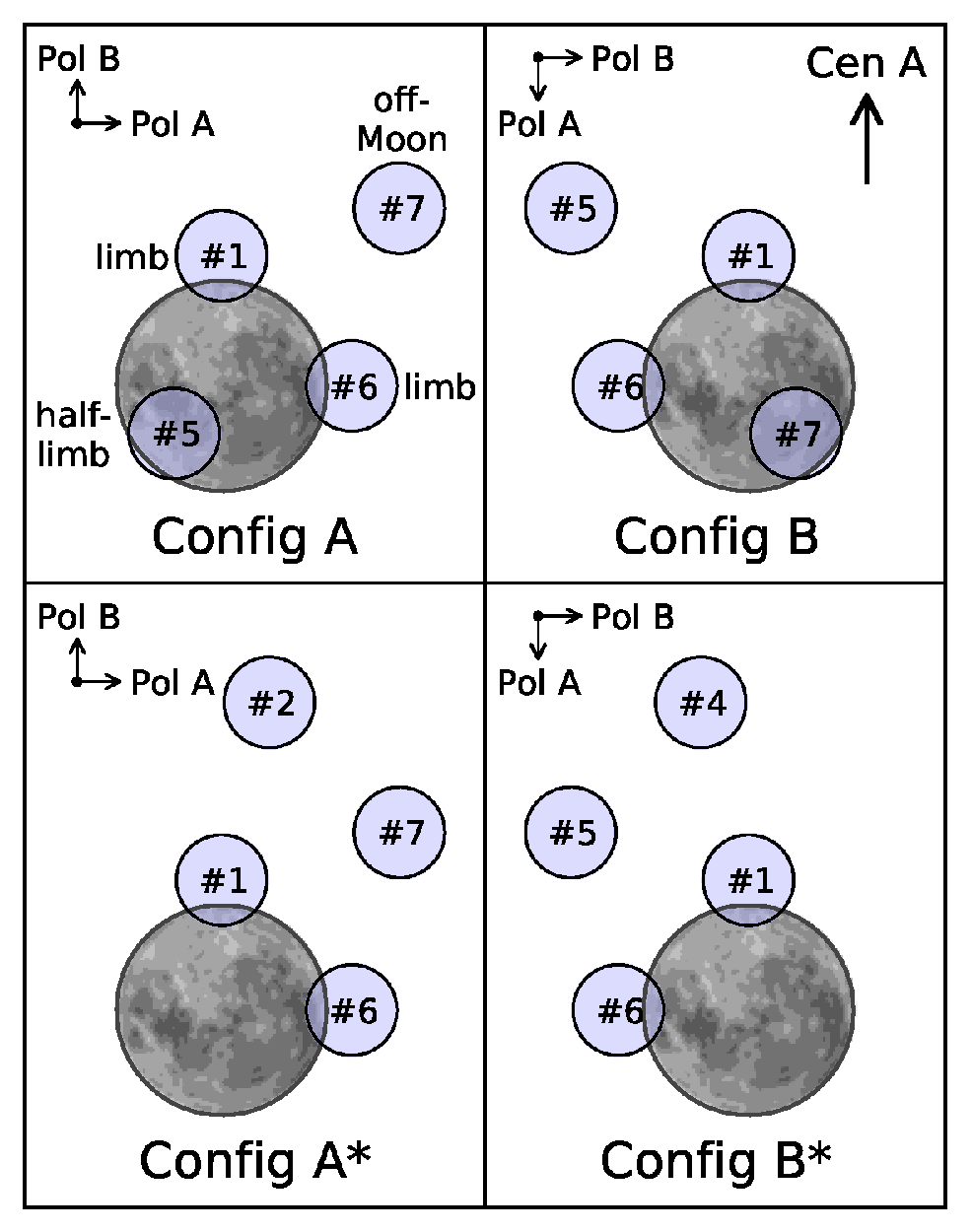}
 \caption{Pointing configurations employed for pulse detection in this experiment; an additional centre-Moon pointing (not shown) was used for calibration, as described in \secref{sec:calibration}.  Each configuration uses a (numbered) subset of the thirteen beams available with the Parkes 21~cm multibeam receiver.  The limb and half-limb beam pointings (labelled, top left) are sensitive to radio pulses from the Moon, while the off-Moon beams are used for anticoincidence filtering.  For each limb beam, one linear polarisation (``Pol~A'' or ``Pol~B'') is aligned radially to the Moon.  In all configurations, the feed was rotated such that \cena\ was towards the top of the figure, as shown; this requirement, and the rotation limit of the feed, required occasional changes between configurations A and B, or A* and B*.  The separation from the lunar centre is 20\arcmin\ for the limb beams and 10\arcmin\ for the half-limb beam.}
 \label{fig:configs}
\end{figure}

Requirements \ref{it:pointreq_limb} and \ref{it:pointreq_moon} are contradictory, but it is possible to partially fulfill both of them by placing a beam slightly off the limb of the Moon.  The layout of the multibeam receiver allows us to meet requirements \ref{it:pointreq_limb}--\ref{it:pointreq_pol} for two beams simultaneously, with both beams being slightly off-limb (20\arcmin\ from the lunar centre, compared to a median lunar radius of 16\arcmin) and having radial/tangential polarisation alignment (see \figref{fig:configs}).  This allows a third beam to be placed on the Moon, although it is in a non-optimal half-limb position (10\arcmin\ from the lunar centre) and has inferior sensitivity.  The fourth beam supported by our backend hardware is placed away from the Moon, and serves to improve the efficacy of our anticoincidence filter.  As it receives no significant contribution from the thermal radiation of the Moon, it is particularly sensitive and thus effective at detecting and identifying local RFI.  Due to the importance of this role, we later changed the configuration to replace the half-limb beam with a second off-Moon beam, as shown for configurations A* and B* in \figref{fig:configs}.

Requirement~\ref{it:pointreq_pol} assumes that the Askaryan pulse maintains its polarisation angle from the Moon to the telescope, neglecting the Faraday rotation it will undergo as it passes through the ionosphere.  The Faraday rotation angle for a thin-layer ionosphere, in radians, is
 \begin{equation}
  \beta = \frac{e^3}{8 \pi^2 m_e^2 \varepsilon_0 c} \, N_e B_{\parallel} \nu^{-2} \label{eqn:faraday}
 \end{equation}
where $\nu$ is the radio frequency, $e$ and $m_e$ the charge and mass of the electron, $\varepsilon_0$ the vacuum permittivity, $c$ the speed of light in vacuum, $N_e$ the column density of free electrons in the atmosphere integrated along the line of sight, and $B_{\parallel}$ the component along the line of sight of the geomagnetic field.  For a typical geomagnetic field of 50~$\mu$T, pessimistically assumed to be parallel to the line of sight, and \mbox{$N_e \sim 10$}--$20 \times 10^{16}$ electrons~m$^{-2}$ (typical for our observations; see \secref{sec:ionosphere}), the maximum rotation angle in our band is 5--10\degree, corresponding to a loss of signal strength of \mbox{$\sim 1$}\% in the expected polarisation.

We applied requirement~\ref{it:pointreq_cena} to the potential UHE particle source \cena, a nearby radio galaxy which is weakly correlated with the arrival directions of detected UHE cosmic rays\gcitep{abreu2010}.  This restricted our observations to a period of 3--5 days in each 27-day lunar cycle during which the Moon makes its closest approach to \cena.  It also required rotation of the feed array to keep one beam at the correct point on the lunar limb.  Due to the feed rotation limit and the motion of the Moon across the sky, it is usually not possible to maintain the required pointing with the same set of beams for an entire observation.  This occasionally made it necessary to switch to a different set of beams, for which the required pointing was within the available feed rotation range.

The required telescope pointing and feed rotation were determined at one-minute intervals with a combination of the \textsc{slalib}\footnote{\url{http://www.starlink.rl.ac.uk/docs/sun67.htx/sun67.html}} and \textsc{PyEphem}\footnote{\url{http://rhodesmill.org/pyephem/}} libraries.  Each pointing was a fixed position on the sky in equatorial coordinates; the motion of the Moon relative to the sky is the dominant source of pointing error, amounting to \mbox{$\pm 0.25$}\arcmin\ within each interval, which is small compared to the beam size.

\subsection{Receiver linearity}
\label{sec:linearity}

For an Askaryan pulse to be detectable by this experiment, it must have a peak amplitude substantially greater than the thermal noise from the Moon, exceeding the dynamic range for which radio telescopes are typically designed.  It is therefore critical to ensure that the receiver, in translating the electric field of the received radio signal to the output signal voltage, maintains a linear response even at large amplitudes.  \citet{jaeger2010} used only antennas of the EVLA in their experiment, despite the availability of additional unupgraded antennas of the VLA (Very Large Array), because they found that the latter failed to maintain the necessary linear response.

\begin{figure}
 \centering
 \includegraphics[width=\linewidth]{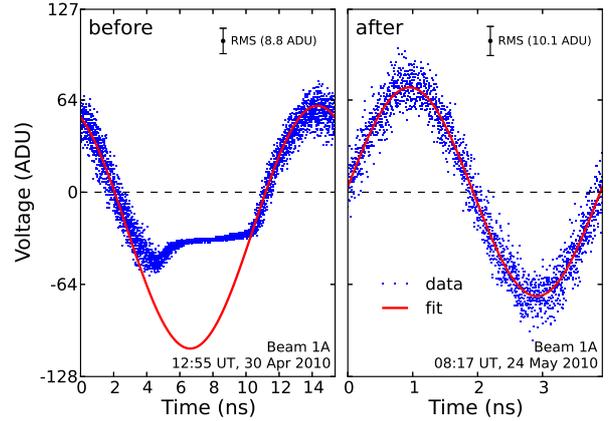}
 \caption{Single buffers of sampled values, recorded while transmitting a sine wave signal directly into the receiver, folded to the period of the signal.  Left and right panels show data from before and after additional attenuation was inserted early in the signal path to restore linear behaviour at large voltage amplitudes.  The solid line shows a sine-wave fit to the data near zero voltage, where the signal behaves linearly in both cases.  The frequency of the signal is different in each case, but we do not expect this to significantly influence the measurement of the non-linearity effect.}
 \label{fig:sinetest}
\end{figure}

For our experiment, we tested the linearity of the response during our initial observations in April 2010, by transmitting a strong monotone signal from a small antenna within the 64~m telescope dish directly into the prime-focus receiver.  Buffers of data were captured via our standard signal path and showed that the receiver was saturating, failing to properly amplify the sine-wave signal in the negative voltage direction (see \figref{fig:sinetest}).  This problem was solved by inserting 20~dB of attenuation prior to the component in the signal path responsible for the non-linear behaviour, returning it to its range of linear operation, and adding an extra 20~dB of gain at a later stage (see \figref{fig:sigpath}).  This hardware change was applied in May 2010 and for all subsequent observations.

\subsection{System dispersion}
\label{sec:system_dispersion}

The detection of a pulse coherent across a range of frequencies, as attempted in this experiment, requires that the components of the pulse at different frequencies arrive simultaneously, to a precision better than the inverse bandwidth.  As conventional radio observations are unaffected by dispersion on this timescale, it is possible that some degree of such dispersion is inherent to the telescope system (not to be confused with ionospheric dispersion; see \secref{sec:ionosphere}).  Previous experiments have measured the impulse response of their receiver systems, but not to the precision required to demonstrate that this effect is negligible.

We tested the system dispersion with a network analyser, measuring the group delay through the receiver system at each frequency within the band, using the same test antenna as in \secref{sec:linearity}.  We found the group delay to have a root mean square (RMS) variation on a 1~MHz scale of 0.15~ns, which is a small fraction of the inverse bandwidth and hence of the expected pulse width.  The resulting amplitude loss relative to an undispersed pulse, calculated by applying corresponding random delays to frequency components of a simulated pulse, is less than $0.01$\%.

\section{Calibration}
\label{sec:calibration}

To establish the sensitivity of our experiment to an Askaryan pulse we must calibrate the response of our instrument, from the spectral electric field strength $\mathcal{E}(\nu)$ of the pulse through to the amplitude of the resulting peak in the signal $s(t)$ recorded in a buffer.  We performed this calibration using the Moon, which is a strong source of thermal emission in our radio band and larger in angular size than our primary telescope beam, separately for each polarisation channel of each beam.  This procedure was more complex than using an unresolved astronomical point source as a calibrator, but served the additional purpose of testing our assumptions about the beam shape and the distribution and polarisation of the lunar thermal emission.

The signal $s(t)$ in a recorded buffer of data from our experiment is measured in analog-to-digital units (ADU).  For a buffer of length $\Delta t$, we find the power spectrum $P(\nu)$ from the signal's Fourier transform $S(\nu)$ as
 \begin{equation}
  P(\nu) = S(\nu)^2 / \Delta t \label{eqn:powspect}
 \end{equation}
where the transform is normalised such that
 \begin{equation}
  \int\! d\nu \, P(\nu) = s_{\rm rms}^2
 \end{equation}
with $s_{\rm rms}$ being the RMS noise level.  The power spectrum is related to the flux density $F(\nu)$ by
 \begin{equation}
  P(\nu) = B(\nu) \, F(\nu) \label{eqn:bandpass}
 \end{equation}
where the bandpass function $B(\nu)$ incorporates the unit conversion.  The flux density contains contributions from the system noise of the receiver and the flux received from a strong source such as the Moon, the latter of which will vary depending on the position of the source in the beam.  For calibration, we pointed each beam successively at the centre of the Moon, in the limb position shown in \figref{fig:configs}, and off the Moon.  In each position, we recorded \mbox{$\sim 1,000$} buffers over the course of a minute, comprising a few ms of data, and calculated averaged power spectra of $P_{\rm moon}(\nu)$, $P_{\rm limb}(\nu)$, and $P_{\rm off}(\nu)$ respectively.  These were subjected to a median filter with width 4~MHz to remove narrow-band interference, and a second-order Savitzky-Golay smoothing filter\gcitep{savitzky1964} with width 16~MHz to reduce the noise level while preserving features of this scale or larger.

\begin{figure}
 \centering
 \includegraphics[width=\linewidth]{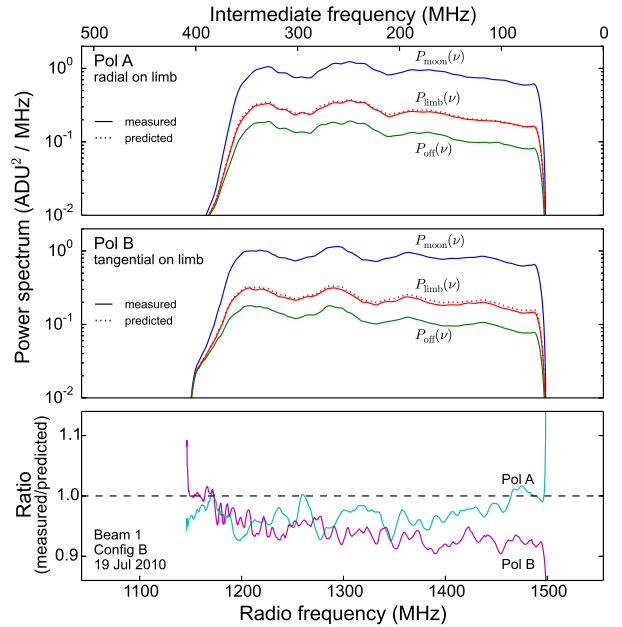}
 \caption{Measured power spectra for centre-Moon, off-Moon and limb pointings, as a function of both RF (bottom axis labels) and IF (top axis labels), for both polarisations (upper and centre panels) of a single beam.  Variation between beams is similar to the variation between polarisations shown here.  For the limb pointing, one polarisation is aligned radially to the Moon, and the other tangentially.  Also shown (dotted) are the spectra predicted for the limb pointings.  Finally, the lower panel shows the ratio between the measured spectra and these predictions.}
 \label{fig:spectcomp}
\end{figure}

The resulting spectra are shown in \figref{fig:spectcomp}.  We note that there is significant structure visible at a scale of \mbox{$\sim 50$}~MHz.  As the thermal spectrum of the Moon contains no such structure, this must result from the bandpass function, which encompasses the effects of all components of the analog portion of the signal path (\figref{fig:sigpath}, left side).  This bandpass structure is quite normal for such wideband analog systems and is quite stable in time.  This frequency structure has no significant effect on the sensitivity of the experiment (see \secref{sec:bandpass}).

Among the analog components of the signal path are the post-mixing attenuators, which were adjusted after each change of pointing configuration to maintain equal power levels in the swapped off- and on-Moon beams.  We therefore performed this calibration routine once per pointing configuration per 3--5 day observing run, except for configuration~A during 24--26 May 2010, which was calibrated based on an assumed equivalence between the received lunar thermal emission in each beam during this run and in the same configuration during the following month; the results are not notably inconsistent.  On one occasion, we calibrated twice on the same day with the Moon at different elevations (56$^\circ$ vs.\ 79$^\circ$) and measured $\sim 2$\% variation in the signal power, which we ascribe to thermal radiation from the ground received through the far sidelobes of the telescope.

The contribution of the Moon to the received flux is found, for a single polarisation, with the Rayleigh-Jeans law:
 \begin{equation}
  F_{\rm moon}(\nu) = \frac{k \nu^2}{c^2} \int\! d\Omega \, \mathcal{B}(\theta, \nu) \, T(\hat{\Omega}, \phi) \label{eqn:rayleigh-jeans}
 \end{equation}
where $k$ is the Boltzmann constant and $\mathcal{B}(\theta, \nu)$ is the beam power pattern, assumed to be radially symmetric (we take it to be an Airy disk) and normalised to \mbox{$\mathcal{B}(0, \nu) = 1$}; $B(\nu)$ is thus the bandpass at the centre of the beam, at \mbox{$\theta = 0$}.  $T(\hat{\Omega}, \phi)$ is the brightness temperature of the Moon at position $\hat{\Omega}$.  As the thermal emission from the Moon is partly linearly polarised, it must be expressed as
 \begin{equation}
  T(\hat{\Omega}, \phi) = \left( T_{\rm tot}(\hat{\Omega}) - T_{\rm pol}(\hat{\Omega}) \right) + T_{\rm pol}(\hat{\Omega}) \cos^2 \phi
 \end{equation}
in terms of the total and polarised components $T_{\rm tot}(\hat{\Omega})$ and $T_{\rm pol}(\hat{\Omega})$, with $\phi$ being the angle between the polarisation of the receiver and the polarisation of the emission.  We take these brightness temperatures from \citet{moffat1972}, who fit a model of the lunar thermal emission to interferometric observations at 1,420~MHz.  Their model incorporates the effects on the intensity and polarisation of the thermal emission from subsurface layers of the Moon as it is refracted through the rough lunar surface, as well as polar cooling.

The true equatorial temperature (i.e.\ the subsurface temperature before the emissivity of the surface is taken into account) in the model of \citet{moffat1972} is 224~K.  This value is approximately constant across a wide band, including the frequencies used in this experiment\gcitep{troitskii1970}.  The model does not incorporate the $\lesssim 5$~K temperature difference between the lunar maria and highland regions, which leads to a $\pm 1$\% uncertainty, nor the variation over a lunar day of about $\pm 3$\% \citep[and references therein]{troitskii1970}.  The systematic error is somewhat larger, with an uncertainty of $\pm 8$\% in the absolute calibration of \citet{moffat1972}, which is typical for radio observations.

With the received flux from the Moon calculated with \eqnref{eqn:rayleigh-jeans}, the received flux from the off-Moon sky assumed to be zero, and measurements of the power spectra $P_{\rm moon}(\nu)$ and $P_{\rm off}(\nu)$, the bandpass function $B(\nu)$ can be calculated.  As a test, we use the calculated bandpass function to determine the expected power spectrum $P_{\rm limb}(\nu)$ in the limb-pointing position, and compare it with the measured spectrum.  A comparison of this type is shown in \figref{fig:spectcomp}; in general, the results are consistent within the range of random error described above.

\begin{figure}
 \centering
 \includegraphics[width=\linewidth]{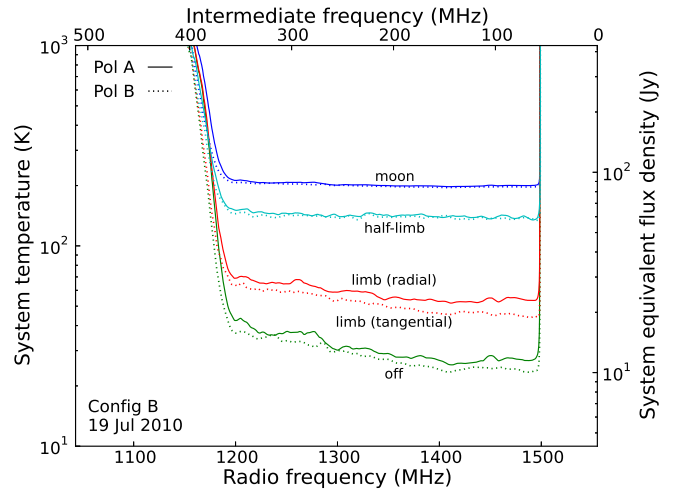}
 \caption{Calibrated spectral noise in terms of the system temperature (left) and the system equivalent flux density (right; for one polarisation only), for four different beam pointings with respect to the Moon.  The limb and half-limb pointings are equivalent to beams \#1 and \#7 respectively in \figref{fig:configs} (configuration~B), and the remaining two cases are for pointing off the Moon (``off'') and at its centre (``moon'').  The 1~Jy offset between the two polarisations for most pointings indicates a difference in the base receiver noise in the two channels; the increased 3~Jy offset for the limb pointing results from the alignment between polarisation~A and the radial polarisation of the lunar thermal emission.}
 \label{fig:tsyscomp}
\end{figure}

We conducted this calibration procedure for each pointing configuration of every observing run; \figref{fig:tsyscomp} shows the calibrated spectral noise in one such case.  The results in each case were internally consistent, but there were differences in the noise level between runs, primarily due to variation in the apparent size of the Moon and hence in the received thermal radiation.  For simplicity in determining the overall sensitivity of this experiment to an Askaryan pulse, we adopt for each distinct beam position and polarisation a single value for the noise, averaging over this variation.  We take the maximum measured variation as the error in the noise power, and hence in the sensitivity, which gives a random uncertainty of $\pm 9\%$, comparable to the $\pm 8\%$ systematic uncertainty.  Both of these dominate over the error in the calibration procedure for each individual run.

The sensitivity to an Askaryan pulse is best expressed as a threshold in the spectral electric field strength immediately prior to being received by the antenna.  As the electric field is proportional to the square root of the power, the uncertainty in this value will be half the uncertainty quoted above.  To relate the electric field strength to the recorded signal $s(t)$, we use its relation\gcitep{james2010} to the flux density for a limited interval $\Delta t$:
 \begin{equation}
  F(\nu) = \frac{1}{Z_0 \Delta t} \, \mathcal{E}(\nu)^2
 \end{equation}
where $Z_0$ is the impedance of free space.  From \eqnrefii{eqn:powspect}{eqn:bandpass} we can then find that
 \begin{equation}
  S(\nu) = \sqrt{ \frac{ B(\nu) }{ Z_0 } } \, \mathcal{E}(\nu) \label{eqn:abscal}
 \end{equation}
for a signal originating from the centre of the beam.  Assuming the signal to comprise a coherent pulse with zero phase at all frequencies (i.e.\ \mbox{$\arg(S(\nu)) = 0$\,;} neglecting the effects described in \secrefii{sec:dedispersion}{sec:phase}), the peak amplitude is
 \begin{equation}
  s_{\rm peak} = \int\! d\nu \, S(\nu) . \label{eqn:fullcoh}
 \end{equation}
Approximating the pulse spectrum $\mathcal{E}(\nu) = \mathcal{E}$ to be constant, and inserting an additional factor $\mathcal{B}(\theta, \nu)$ to allow for a signal originating away from the centre of the beam, \eqnrefii{eqn:abscal}{eqn:fullcoh} then give us
 \begin{equation}
  \mathcal{E} = s_{\rm peak} \left( \int\! d\nu \, \sqrt{ \frac{ B(\nu) \mathcal{B}(\theta, \nu) }{ Z_0 } } \right)^{-1} \label{eqn:thresh}
 \end{equation}
as the threshold spectral electric field strength for the detection of such a pulse.  Note that $s_{\rm peak}$ may also be represented as
 \begin{equation}
  s_{\rm peak} = n_\sigma s_{\rm rms} \label{eqn:nsigma}
 \end{equation}
in terms of the significance $n_\sigma$ of the pulse relative to the RMS noise.  The value of this threshold significance is found through analysis of the observational data in \secref{sec:obs}.

\section{Ionospheric dispersion}
\label{sec:ionosphere}

A major consideration for this type of experiment is the dispersion of coherent pulses as they pass through the ionosphere, which causes a frequency-dependent delay in the pulse arrival time.  For a wide bandwidth, the effect is to stretch the pulse out in time and to reduce its peak amplitude, particularly at low frequencies.  Early experiments looked in multiple frequency bands for coincident pulses, separated by an offset corresponding to the dispersive delay\gcitep{hankins1996,beresnyak2005}.  Others, operating at high frequencies with relatively narrow bandwidths, simply neglected the minor dispersive effects\gcitep{gorham2004a,jaeger2010}.  \citet{buitink2010}, whose experiment was the most severely affected by dispersion due to its lower frequency, recorded voltage data continuously over the course of their observations, and dedispersed it afterwards to restore the original pulse.  \citet{james2010} used an analog dedispersion filter developed by \citet{roberts2008} which allowed the same correction to be performed in real time, but it was limited to a fixed dispersion characteristic, whereas the effect of the ionosphere varied over the course of the observations.

For this experiment, we developed a digital dedispersion filter (see \secref{sec:dedispersion}) which could be adjusted to compensate for different degrees of dispersion in real time.  This allowed us to detect a dispersed pulse and trigger the storage of buffered data, so that more precise corrections could be applied in retrospective processing.  To determine the correct setting for the filter, we required real-time information about the degree of dispersion, which is determined by the ionospheric total electron content (TEC).  This is typically reported as a column number density in TEC units ($1~{\rm TECU} = 10^{16}\ {\rm electrons~m}^{-2}$), either for a vertical column (vertical TEC, or VTEC) or a column along the line of sight (slant TEC, or STEC).

\begin{figure}
 \centering
 \includegraphics[width=\linewidth]{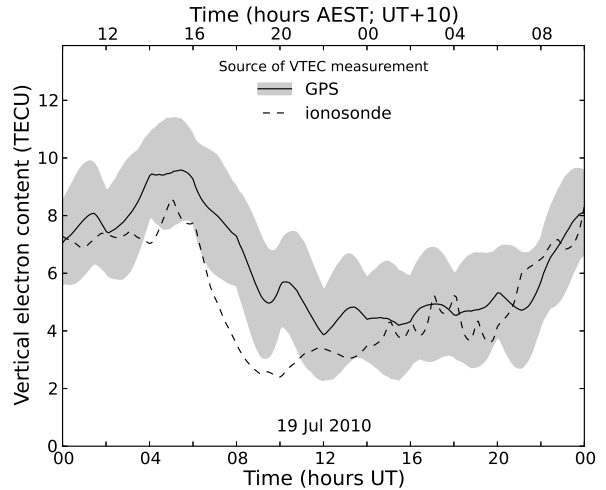}
 \caption{GPS- and ionosonde-derived VTEC over a single day at the Parkes telescope.  The grey shading represents the uncertainty in the GPS measurement.  No uncertainty bounds are available for the ionosonde data, but its deviation from the GPS values provides an indication of its precision.}
 \label{fig:parkes_vtec_short}
\end{figure}

\begin{figure}
 \centering
 \includegraphics[width=\linewidth]{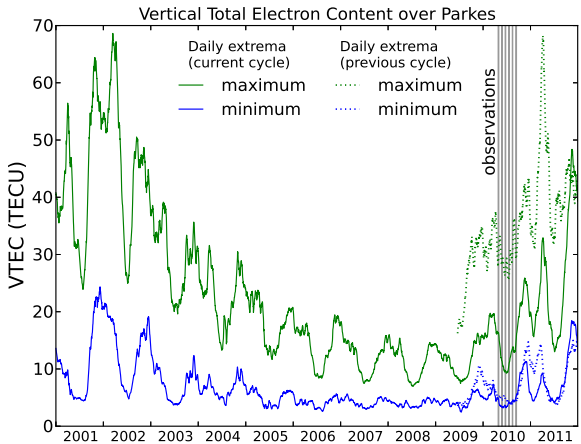}
 \caption{Electron content of the ionosphere over Parkes, derived from \gls{GPS} data.  Values shown are the maximum and minimum per 1-day cycle, smoothed with a width of 20~days, for the most recent 11-year solar cycle and the tail end of the previous one.   At the time of our observations (vertical bars) the electron content was quite low, particularly compared to the equivalent point in the previous solar cycle.}
 \label{fig:parkes_vtec_long}
\end{figure}

The free electron content of the ionosphere is controlled by ionisation caused by the high-energy component of solar radiation, and subsequent recombination with positive ions over time.  As a result, it varies with latitude and season, and peaks several hours after local noon (see \figref{fig:parkes_vtec_short}).  It also varies with the 11-year solar magnetic activity cycle; our observations were conducted towards the end of the most recent solar minimum (see \figref{fig:parkes_vtec_long}).

For our experiment, we used TEC values derived from two different sources.  The first of these is the Global Positioning System (GPS), which measures the dispersion of radio signals between a network of ground stations and a constellation of satellites in medium Earth orbit.  These values are the most accurate, but were only available retrospectively, after a delay of several weeks.  The second source is ionosonde measurements, which use active radar to probe the reflectivity of the plasma in the lower ionosphere.  These values are less accurate, and neglect conditions in the upper ionosphere, but were available in near real-time.

\subsection{GPS electron content measurements}
\label{sec:gps}

Dual-frequency GPS measurements determine the STEC along the lines of sight between satellites and ground-based receivers to a precision of better than 0.1~TECU\gcitep{hernandez-pajares2009}.  These STEC measurements at a sparse set of points, between each receiver and each visible satellite, are then interpolated to produce a global map of ionospheric VTEC by several groups; we used the maps generated by the Center for Orbit Determination in Europe (CODE).  However, this procedure introduces substantial uncertainty, due both to the interpolation between sparse measurements and the assumption that the ionosphere can be represented with a single VTEC value at each coordinate, implying that it exists as a single thin layer.  CODE also supplies maps of the estimated uncertainty in the VTEC measurement; during our observations, this was typically $\pm 2$~TECU (see \figref{fig:parkes_vtec_short}).

\begin{figure}
 \centering
 \includegraphics[width=0.6\linewidth]{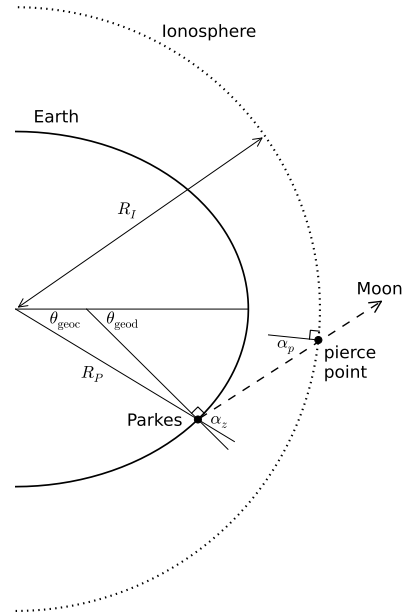}
 \caption{Determination of the ionospheric pierce point on the line of sight (dashed) from the Parkes telescope to the Moon, for the single-layer ionosphere model.  Note the distinction between geodetic latitude $\theta_{\rm geod}$ and geocentric latitude $\theta_{\rm geoc}$; the zenith angle $\alpha_z$ is defined with respect to the geocentric vertical.  The ratio between \gls{VTEC} and \gls{STEC} is determined by the angle of incidence $\alpha_p$ of the line of sight at the pierce point, as in \eqnref{eqn:slant}.  (Not to scale.  The oblateness of the Earth has been exaggerated).}
 \label{fig:ionomodel}
\end{figure}

The VTEC maps were obtained from the Crustal Dynamics Data Information System (CDDIS)\footnote{\url{ftp://cddis.gsfc.nasa.gov/pub/gps/products/ionex/}}\gcitep{noll2010}, in the IONEX (IONosphere map EXchange) format\gcitep{schaer1998}.  A single file in this format specifies the VTEC on a regular grid in geocentric latitude/longitude (i.e.\ on a spherical shell) at intervals over the course of a day.  To find the STEC during our observations, we first determined the pierce point on this shell of the line of sight between the Parkes telescope and the Moon as shown in \figref{fig:ionomodel}, and performed bilinear interpolation between spatially adjacent grid points to find the VTEC.  We interpolated in time between the previous and next VTEC maps, rotating them in longitude to match the local time at the pierce point, using the third and most precise interpolation method described by \citet{schaer1998}.  Finally, we converted from VTEC to STEC using a slant factor based on the angle $\alpha_p$ between the line of sight and the normal to the ionosphere at the pierce point, given by
 \begin{align}
  \frac{\rm \gls{STEC}}{\rm \gls{VTEC}} &= \frac{1}{\cos\alpha_p} \\
   &= \left( 1 - \left( \frac{R_P}{R_I} \sin\alpha_z \right)^2 \right)^{-1/2} \label{eqn:slant}
 \end{align}
where $\alpha_z$ is the zenith angle relative to the geocentric vertical, and $R_P$ and $R_I$ are the radii from the centre of the Earth of the Parkes telescope and the model ionosphere shell respectively.  The \gls{CODE} maps assume a single-layer ionosphere at an altitude of 450~km relative to a mean Earth radius of 6,378~km, for a total $R_I = 6$,828~km.

The error introduced by the above procedure for determining the STEC from a CODE VTEC map should be substantially smaller than the map's reported uncertainty, so we display only the latter uncertainty elsewhere in this paper, scaled by the slant factor.  This uncertainty was typically \mbox{$\pm 2$} TECU, with a maximum value of 3.8~TECU over the course of our observations.  These values omit the dispersive effect of electron content at altitudes exceeding that of GPS satellites (20,200~km) but, applying a model of the high-altitude electron distribution\gcitep{gallagher1988}, we find this to be \mbox{$\lesssim 0.1$} TECU, which is negligible.

\subsection{Ionosonde electron content measurements}
\label{sec:ionosonde}

An ionosonde is a radar used to probe conditions in the ionosphere, typically operating in the frequency range 1--20~MHz.  The maximum frequency at which it detects a reflection from the F2 layer of the ionosphere, referred to as foF2, is a measure of this layer's plasma frequency and hence of its electron density.  The Ionospheric Prediction Service (IPS), a branch of the Australian Bureau of Meteorology, makes available IONEX-format maps of VTEC based on foF2 measurements from a network of ionosondes\footnote{\url{ftp://ftp.ips.gov.au/data/Satellite/}}.

As foF2 directly measures only the peak electron density in the F2 layer, rather than the integrated column electron density, ionosonde data are less accurate than GPS-derived TEC values.  \Figref{fig:parkes_vtec_short} shows the variation between data from these two sources.  The advantage of the ionosonde VTEC maps, however, is that they are available in near real-time, with a delay of only 1--2 hours.  This allowed us to use them to calibrate our dedispersion filter during our observations.  We derived STEC values from VTEC maps as for GPS data in \secref{sec:gps}, and estimated the current value as
 \begin{equation}
  {\rm STEC}_{\rm est}(t) = {\rm STEC}(t - \Delta t) \times \frac{ {\rm STEC}(t - 24\,{\rm hrs}) }{ {\rm STEC}(t - 24\,{\rm hrs} - \Delta t) } ,
 \end{equation}
scaling the most recent available value (at time \mbox{$t - \Delta t$}) by the change in STEC during the equivalent period on the previous day.

\begin{figure}
 \centering
 \includegraphics[width=\linewidth]{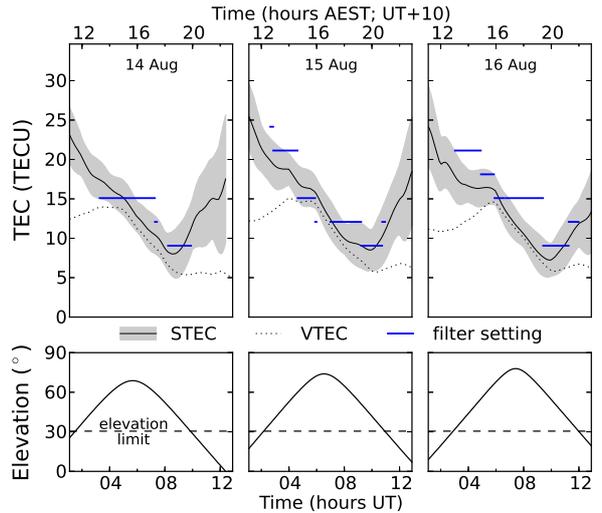}
 \caption{Comparison between ionospheric TEC and the STEC for which our dedispersion filter was set during our three-day observing run in August 2010.  The upper plots show the GPS-derived STEC along the line of sight to the Moon (solid; uncertainty range shaded) and VTEC (dotted), as well as the filter settings (horizontal bars) with which we attempted to match the STEC.  Ultimately, the filter settings were sufficiently closely matched to the STEC that there was no detriment to the retrospective sensitivity of the experiment (see \secref{sec:threshold}).  The lower plots show the elevation of the Moon relative to the elevation limit of the Parkes radio telescope; the increased STEC relative to the VTEC at the start and end of each observation is due to the Moon being close to the horizon.}
 \label{fig:tec_obs}
\end{figure}

As no error estimate is available for the ionosonde-derived TEC values, we take as the error the difference between them and the more accurate values derived from GPS measurements, which was typically \mbox{$\pm 4$} TECU in STEC.  During our observations, we set the dedispersion filter to correspond to the multiple of 3.02~TECU (corresponding to 1~ns of dispersion across the RF band 1.2--1.5~GHz; see \secref{sec:dedispersion}) closest to the ionosonde-derived STEC value.  The resulting variation between the filter setting and the GPS-derived STEC is shown in \figref{fig:tec_obs} for a typical observing run; the maximum value of this error over the course of our observations was 9.0~TECU.  We describe how we dedispersed the signal in \secref{sec:dedispersion}, and the consequences of the STEC error and uncertainty for the sensitivity of our experiment in \secref{sec:signal_sens}.

\section{Signal optimisation}
\label{sec:signal}

The signal-processing problem for an experiment of this type is to detect an Askaryan pulse against a Gaussian noise background dominated by thermal radiation from the Moon and internal noise in the telescope receiver (apart from RFI, which we address in \secref{sec:cuts}).  Previous experiments have either converted the signal to the power domain or searched directly in the voltage domain, using a range of filters to improve the signal-to-noise ratio.  Here we use the voltage-domain approach, using the theoretically optimum\citep[e.g.][Ch.\ 42]{byrne2005} filtering strategy: the maximum signal-to-noise ratio is achieved by applying a pre-whitening filter with a bandpass that gives the noise a flat (white) spectrum, followed by a filter matched to the expected shape of the pulse.  In terms of their respective transfer functions $W(\nu)$ and $M(\nu)$, we can represent these as a single combined filter with transfer function
 \begin{equation}
  O(\nu) = W(\nu) M(\nu) .
 \end{equation}
We divide our discussion of the optimising filter $O(\nu)$ into three parts.  In \secref{sec:bandpass}, we describe the derivation of its amplitude $|O(\nu)|$, which represents the optimum bandpass of the filter.  In \secref{sec:dedispersion}, we consider the derivative of the phase $\arg(O(\nu))$ across the band: the first-order term or linear phase slope corresponds to an absolute delay in the arrival time of a pulse, which does not affect its detectability, but the second- and higher-order terms describe the dispersion resulting from the passage of the pulse through the ionosphere (see \secref{sec:ionosphere}), which is reversed by an appropriate filter.  In \secref{sec:phase}, we consider the base phase $\arg(O(0))$ of the filter, which relates to the inherent shape of the pulse.

The optimising filter described above maximises the pulse amplitude relative to the background noise for an analog signal.  In the digital domain, in which the filter is implemented, the signal is represented by a series of discrete samples and must therefore be interpolated to restore the full pulse amplitude, as described in \secref{sec:interp}.  In our experiment, both the filter and this interpolation were applied in an approximate form in real time, and more precisely in retrospective processing.  In \secref{sec:signal_sens} we discuss the signal loss in each case from the combination of the above effects.

\subsection{Bandpass optimisation}
\label{sec:bandpass}

Prior to filtering, the signal has the noise power spectrum $P(\nu)$, from \eqnref{eqn:powspect}.  To give the signal a flat noise spectrum, the pre-whitening filter must have a bandpass
 \begin{equation}
  |W(\nu)|^2 \propto 1 / P(\nu) .
 \end{equation}
The purpose of this filter is to place a greater weight on the signal at frequencies with lower noise power, at which the signal-to-noise ratio is higher and the system is therefore more sensitive.

The pre-whitening filter should optimally be followed by a matched filter with transfer function $M(\nu)$, with its bandpass $|M(\nu)|^2$ matched to the spectrum of the expected pulse.  If the pulse initially has an electric field spectrum $\mathcal{E}(\nu)$, and hence a power spectrum $\mathcal{E}(\nu)^2$, it will be modified first by the bandpass $B(\nu)$ of the receiver, and then the bandpass $|W(\nu)|^2$ of the pre-whitening filter, so
 \begin{align}
  |M(\nu)|^2
   &\propto |W(\nu)|^2 B(\nu) \mathcal{E}(\nu)^2 \\
   &\propto \frac{ \mathcal{E}(\nu)^2 B(\nu) }{ P(\nu) } .
\end{align}
The bandpass of the combined optimising filter is then
 \begin{align}
  |O(\nu)|^2
   &\propto (|W(\nu)||M(\nu)|)^2 \\
   &\propto \frac{ \mathcal{E}(\nu)^2 B(\nu) }{ P(\nu)^2 }
 \end{align}
which gives us the amplitude of its transfer function $O(\nu)$.

We applied this correction in retrospective processing separately for each beam, using the bandpass $B(\nu)$ derived in \secref{sec:calibration}, and taking $\mathcal{E}(\nu)$ to be the electric field spectrum for a cascade observed in the most optimistic case, from the Cherenkov angle, in order to minimise the threshold detectable particle energy\gcitep{alvarez-muniz2006}.  Typical improvements in the signal-to-noise ratio were 1--2\%, and not sensitive to our assumption regarding the original pulse spectrum, resulting instead from corrections to minor variations in the bandpass and noise spectrum.  This improvement is insignificant compared to the calibration uncertainty (see \secref{sec:calibration}), indicating that these variations (see \figref{fig:spectcomp}) have a negligible effect on the sensitivity, and we neglect this improvement in subsequent calculations.

The effect of bandpass optimisation will be greater for future experiments with larger fractional bandwidths.  The significance of the assumption regarding the pulse spectrum will also increase, and it may become necessary to choose between optimising for maximum sensitivity to less energetic particles, as we have here; or optimising for maximum aperture to more energetic particles, by biasing the filter towards low frequencies.  This latter possibility is a more general case of a result found by \citet{james2009b} under the assumption of a flat bandpass: that the aperture of a lunar radio experiment is improved under some circumstances by simply excluding the high-frequency end of its observing band.

\subsection{Dedispersion}
\label{sec:dedispersion}

For radio waves significantly above the plasma frequency ($\nu_p \lesssim 10$~MHz in the ionosphere), the dispersive time delay relative to a signal traveling at the speed of light is
 \begin{align}
  \Delta t &= \frac{ e^2 }{ 8 \pi^2 \varepsilon_0 m_e c } N_e \nu^{-2} \label{eqn:dispersion} \\
           &= 1.34 \times 10^{-7} \left( \frac{ N_e }{{\rm m}^{-2}} \right) \left( \frac{\nu}{\rm Hz} \right)^{-2} {\rm s} .
 \end{align}
Note the similarity of \eqnref{eqn:dispersion} to the expression for Faraday rotation, another ionospheric effect, in \eqnref{eqn:faraday}: both are linearly dependent on $N_e$, the ionospheric STEC.  The delay $\Delta t$ is equivalent to multiplying the signal, in the frequency domain, by a phase factor $e^{i \phi}$ where
 \begin{align}
  \phi &= - 2 \pi \int_\infty^\nu \! d\nu \, \Delta t \label{eqn:dispersion_phase}
 \end{align}
is the integral from an infinite frequency at which \mbox{$\Delta t = 0$}, though the choice of this limit affects only the base phase, not the dispersion.  The delay can be reversed, and the signal dedispersed, by applying the conjugate phase factor $e^{-i \phi}$.  If the signal is shifted in frequency prior to dedispersion, as in this experiment, the frequency $\nu$ in \eqnref{eqn:dispersion} represents the original radio frequency at which dispersion occurs, while the integral in \eqnref{eqn:dispersion_phase} is instead performed over the intermediate frequency at which the dedispersion filter is applied.

\subsubsection{Filter design}
\label{sec:filt_design}

To perform this correction in full, a filter should transform the signal to the frequency domain, multiply by the appropriate phase factor, and transform the result back to the time domain, and we did this in retrospective processing.  However, this was computationally impractical for us to achieve in the several $\mu$s available in real time, to detect a pulse while it was still present in the buffered data.  The filter shown in \figref{fig:sigpath} implements a convolution of the digital sampled data, in the time domain, with 64 arbitrary coefficients which describe its impulse response; in engineering parlance, it is a 64-tap finite impulse response (FIR) filter, with a delay between taps equal to the sampling interval of 0.98~ns.  For this filter to act as an adjustable dedispersion filter, we needed to precompute and store an appropriate set of coefficients for each amount of dedispersion it was to apply.

To determine the optimum filter coefficients, we first calculated the required phase factor at a set of discrete frequencies separated by $\Delta\nu = 0.5$~MHz, and applied an inverse discrete Fourier transform to find the corresponding impulse response.  The length of this impulse response is $1 / \Delta\nu = 2$~$\mu$s, or \mbox{$\sim 2$,000} taps.  As our filter only supports 64 taps, we truncated the impulse response to this length, selecting the set of 64 consecutive coefficients which maximised the retained power (i.e.\ the sum of the squares of the selected values), which we term the filter efficiency.  The filter efficiency decreased significantly when calculating coefficients for a greater degree of dedispersion (\mbox{$\gtrsim 100$} TECU), for which the required impulse response extends over a longer time and exceeds the 64-tap limit.

\begin{figure}
 \centering
 \includegraphics[width=\linewidth]{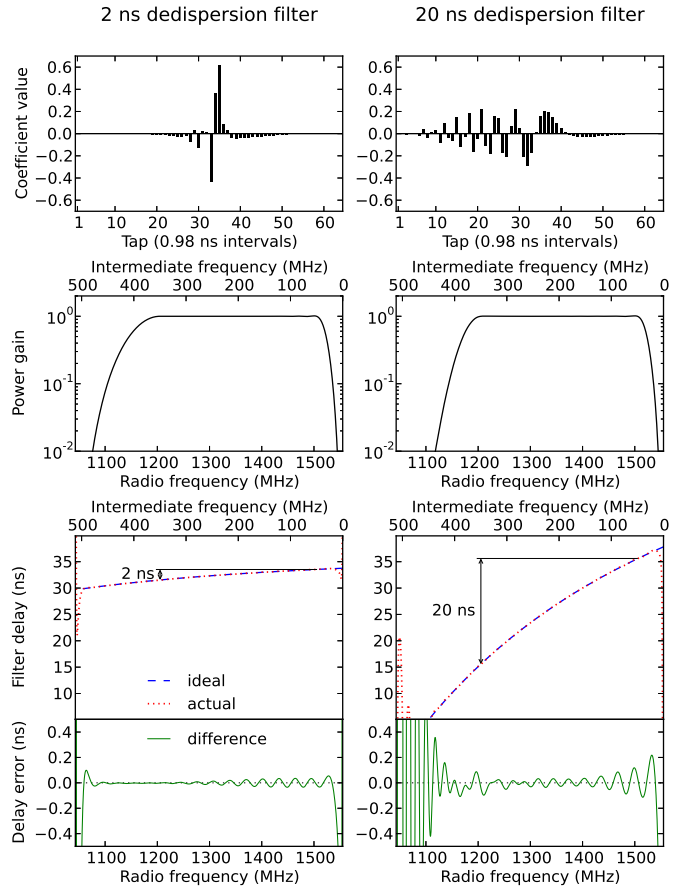}
 \caption{Performance characteristics of our real-time dedispersion filter when set to compensate for dispersion of 2~ns (left) and 20~ns (right) across the 1.2--1.5~GHz radio frequency range of our receiver.  In the impulse response (top), there is a visible separation of the rapidly-oscillating high-frequency components (tap \mbox{$\lesssim 32$}) and the low-frequency components (tap \mbox{$\gtrsim 32$}): this corresponds to a greater group delay at high radio frequencies, counteracting the greater delay at low radio frequencies from ionospheric dispersion.  The ideal (dashed) and actual (dotted) values of the group delay (bottom) correspond closely within the useful band; the deviation between the two (solid) is greater for the 20~ns filter than for the 2~ns filter, but is in either case typically \mbox{$< 0.1$}~ns.  The absolute value of the delay is arbitrary, and corresponds roughly to half the length of the filter.  The periodic oscillation in the delay error plot is similar to the behaviour of the analog filter developed by \citet{roberts2008}.}
 \label{fig:filts}
\end{figure}

Using the above procedure, we generated sets of filter coefficients corresponding to dispersive delays of multiples of 1~ns across the radio frequency range 1.2--1.5~GHz (equivalent to multiples of 3.02~TECU in STEC), allowing us to recalibrate the filter during the experiment by loading new sets of coefficients.  \Figref{fig:filts} shows the impulse response and bandpass of two such sets of coefficients, as well as the accuracy of the resulting dedispersion.  The filter performance declined for dispersive delays greater than \mbox{$\sim 30$}~ns, due to loss of efficiency from truncation of the impulse response.  However, ionospheric TEC during our observations was significantly lower than expected from the previous solar activity cycle (see \figref{fig:parkes_vtec_long}), so the maximum filter setting used in our experiment was only 8~ns, for which the filter efficiency was \mbox{$> 99.9$}\%.  The signal loss in dedispersion is instead dominated by uncertainty in the ionospheric TEC, as discussed in \secref{sec:signal_sens}.

\subsection{Pulse phase}
\label{sec:phase}

Previous experiments of this type, when searching for a pulse in the voltage signal, have searched for an excursion in the voltage beyond some threshold magnitude.  This implicitly makes the same assumption as in \eqnref{eqn:fullcoh}: that the signal has a spectrum $S(\nu)$ with zero phase, so in the time domain it combines coherently in a single direction.  Experimental tests of the Askaryan effect show that this is not the case: the phase is closer to \mbox{$-90$\degree}, and it manifests in the time domain as a bipolar pulse\gcitep{miocinovic2006}, consistent with more recent simulations\gcitep{alvarez-muniz2010}.  In theoretical terms, the two poles of the pulse can be understood as originating from the beginning and end of the particle cascade; the order in which they are observed depends on whether the observer is on the inside or outside of the Cherenkov cone.  More precisely, one pole represents the increase in excess negative charge of the cascade, and the other represents its decrease: the pulse profile is the derivative of the charge excess along the length of the cascade\gcitep{alvarez-muniz2000}.  The practical effect is that the power of the pulse, rather than being concentrated into a single unidirectional excursion of the electric field, is split between the two poles, and its maximum amplitude is decreased by a factor \mbox{$\sim \sqrt{2}$}.

\begin{figure}
 \centering
 \includegraphics[width=\linewidth]{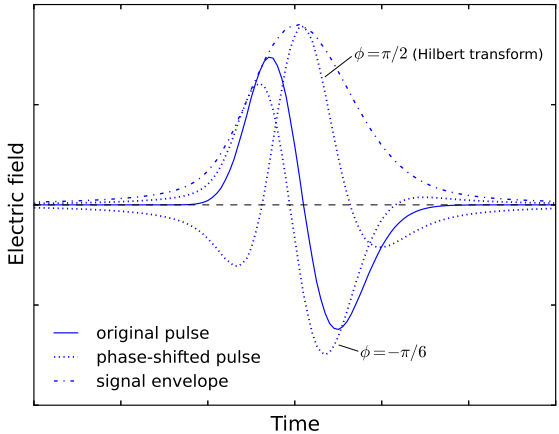}
 \caption{An idealised Askaryan pulse (solid), represented here as the derivative of a Gaisser-Hillas function\gcitep{gaisser1977}.  Applying an arbitrary phase $\phi$ results in a range of other possible pulse shapes, two of which are shown (dotted).  The signal envelope (dash-dotted) reconstructs the maximum possible amplitude of the pulse.}
 \label{fig:phase}
\end{figure}

A filter matched to the expected pulse would apply a phase of 90\degree\ (i.e.\ a Hilbert transform).  However, in our experiment the phase is randomised when the RF signal is downconverted to IF by mixing it with a local oscillator of unknown phase\gcitep{bray2012}, and the resulting pulse will have one of a range of possible profiles (see \figref{fig:phase}).  Instead, in retrospective processing of the data, we reconstruct the maximum possible amplitude of the pulse using the signal envelope as defined by \citet{longuet-higgins1957}.  A side effect of this procedure is to increase the noise level, as the signal envelope has a Rayleigh rather than a Gaussian distribution of amplitude, which must be taken into account when determining the significance of a pulse.

\subsection{Interpolation}
\label{sec:interp}

The digital representation of a signal consists of a series of values corresponding to the voltage of the original analog signal at evenly-spaced sampling times.  \citet{james2010} note that, for a finite sampling rate, there will be a random offset between the peak of a pulse and the closest sampling time, which reduces the sensitivity of an experiment of this type, as the sampled digital data will not record the full magnitude of the pulse.  They calculate the loss of sensitivity this effect causes for their experiment, finding it to be \mbox{$\sim 0$}--30\%, depending on the actual offset between the peak of the pulse and the closest sampling time, and on the coherency of the pulse; other similar experiments have neglected this effect entirely.

In this experiment, we correct for this effect by interpolating (up-sampling) the digital data, reconstructing the values of the analog signal at the times which were not originally sampled.  Perfect interpolation requires a convolution of the signal with a $\sinc$ function to reconstruct the signal at every intermediate point.  The real-time hardware used in our experiment convolves the signal with a digital approximation to a $\sinc$ function, interpolating a single value between every pair of samples.  This doubles the effective sampling rate, halving the expected offset between the peak and the closest sampling time.  Approximating the pulse as a $\sinc$ function, which has quadratic behaviour around its peak, the effect of this is to decrease the loss of sensitivity relative to the case of perfect interpolation by a factor of four.  For retrospective processing, we performed 32-fold interpolation in software, with the remaining loss of sensitivity being negligible (reduced by a factor $32^2$).

\subsection{Signal loss}
\label{sec:signal_sens}

\begin{figure}
 \centering
 \includegraphics[width=\linewidth]{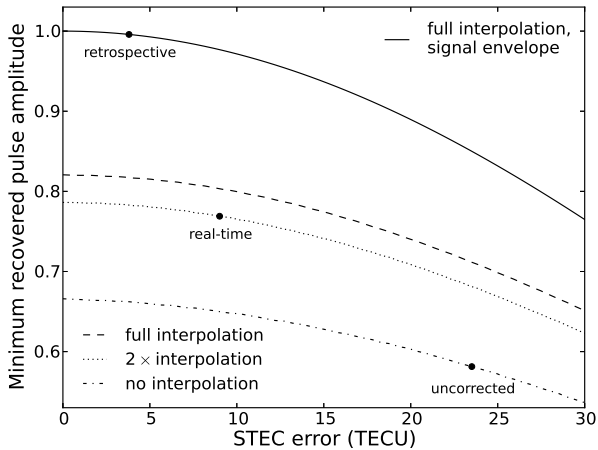}
 \caption{Maximum loss of sensitivity from suboptimal signal-processing in our experiment.  In all cases, the sensitivity decreases if there is an error between the ionospheric STEC assumed by the dedispersion filter and the actual dispersion of the pulse.  The solid line shows the recovered pulse amplitude after the full retrospective processing applied to recorded data, in which we perform effectively complete interpolation and form the signal envelope.  The point labelled ``retrospective'' marks the value based on the maximum uncertainty of the GPS-derived STEC used in this processing.  The dashed line shows the minimum recovered pulse amplitude if we had neglected to form the signal envelope, for the worst-case signal phase.  The dotted line shows the minimum recovered pulse amplitude in real-time processing, in which only two-fold interpolation was performed.  The point labelled ``real-time'' marks the value for the worst-case error in the dedispersion filter setting, equal to the maximum difference between the filter setting and the GPS-derived STEC.  The dash-dotted line shows the minimum recovered pulse amplitude with no processing apart from dedispersion, for the worst-case signal phase and sampling times.  The point labelled ``uncorrected'' marks the value with the absence of any dedispersion, for the maximum STEC during our observations.}
 \label{fig:sensloss}
\end{figure}

In our experiment, in real time, we applied a filter which dedisperses the signal (see \secref{sec:filt_design}), manually adjusting the filter setting based on ionosonde-derived STEC values (see \secref{sec:ionosonde}), and performed two-fold interpolation to double the effective sampling rate.  In retrospective processing, we applied full bandpass optimisation and more precise dedispersion based on GPS-derived STEC values (see \secref{sec:gps}), formed the signal envelope to search through possible phases of the signal, and performed effectively complete (32-fold) interpolation.  To determine the potential loss of signal strength in each case, we simulated the progress of test pulses with different degrees of dispersion through the corresponding signal path, with the results shown in \figref{fig:sensloss}.  The maximum loss of signal strength corresponds to the worst-case values for the signal phase and sampling times.  Due to its minimal effect, we omitted bandpass optimisation from our simulation.  Noise was omitted from our simulation: since the noise has no inherent phase or dispersion, it is equally valid to consider the effects of noise before or after processing.

The effects described in \secrefs{sec:dedispersion}{sec:interp} are not clearly separable in their influence on the loss of signal strength, but we can give some indication of their relative significance by describing the magnitude of each of them individually, with the other effects completely removed.  If the sampled digital data in this experiment had been subject to no processing other than a direct threshold test, the maximum signal loss would have been 17.9\% due to the unknown pulse phase, 21.6\% due to the finite sampling rate, or 15.0\% due to dispersion for the maximum STEC of 23.5~TECU during the entire observations; the maximum total signal loss would have been 41.9\% from these three effects combined (\figref{fig:sensloss}, ``uncorrected'').  With the actual processing implemented in real time, the loss due to the unknown pulse phase remains 17.9\%, but two-fold interpolation reduces the sampling loss to 5.6\%, and the dedispersion filter reduced the dispersive loss to 2.3\%, calculated based on the maximum difference of 9.0~TECU between the filter setting and the STEC values retrospectively derived from GPS data; the total loss from these effects is 23.1\% (\figref{fig:sensloss}, ``real-time'').  In retrospective processing of recorded data, the potential losses due to pulse phase and sampling are effectively eliminated by forming and interpolating the signal envelope, so the only remaining loss is 0.4\% due to dispersion, calculated based on the maximum uncertainty of 3.8~TECU in the GPS-derived STEC values used for retroactive dedispersion (\figref{fig:sensloss}, ``retrospective'').  The relation of these signal losses to the final sensitivity of this experiment is discussed in \secref{sec:threshold}.

\section{Observations and analysis}
\label{sec:obs}

Our observations were conducted from April to September 2010 in six runs of 3--5~days each month, using the observing strategy outlined in \secref{sec:design} and the calibration procedure described in \secref{sec:calibration}.  From 259.5~hours of scheduled time with the Parkes radio telescope, we obtained 148.7~hours of calibrated on-source data.  Details of our observation schedule are in \tabref{tab:obs}.  The length of the buffer recorded on a successful trigger was set to 4~$\mu$s, except for the May observing run, during which it was set to 2~$\mu$s.

\begin{table*}
 \centering
 \begin{threeparttable}
  \caption{Details of our observations in 2010, including pointing configurations (see \figref{fig:configs}) and numbers of triggers in on-Moon beams.}
  \begin{tabular}{lccrcrrcrr}
 \toprule
 \multirow{2}{*}{Date} & \multicolumn{2}{c}{Time (AEST)} & \multicolumn{1}{@{}c@{}}{ \multirow{2}{*}{ \specialcell{r}{Duration\,\tnote{a}\\(hrs)\hphantom{\,\tnote{a}}}\hspace{-\tnotealen} } } & \multirow{2}{*}{Configs} & \multicolumn{2}{c}{Real-time triggers} & & \multicolumn{2}{c}{Triggers over $8\sigma$} \\
 \cmidrule{2-3} \cmidrule{6-7} \cmidrule{9-10}
  & start & end & & & raw & accepted & & accepted & passed cuts \\
 \midrule
 27 Apr\,\tnote{b} & 23:32 & 02:36 & 3.1 & B & 179\,017 & 37\,563\,\tnote{c}\hspace{-\tnoteclen} & & 12 & 4 \\
 28 Apr\,\tnote{b} & 20:23 & 04:27 & 7.1 & B & 292\,925 & 18\,952\,\tnote{c}\hspace{-\tnoteclen} & & 9 & 0 \\
 29 Apr\,\tnote{b} & 21:30 & 05:27 & 7.5 & B & 16\,075\,833 & 37\,538 & & 112 & 5 \\
 30 Apr\,\tnote{b} & 22:00 & 06:33 & 7.1 & B & 16\,288\,167 & 64\,257 & & 78 & 0 \\
 \midrule
 24 May & 18:51 & 01:15 & 5.8 & A,\,B & 17\,042\,475\,\tnote{d}\hspace{-\tnotedlen} & 35\,898 & & 34 & 6 \\
 25 May & 18:12 & 02:09 & 7.5 & A,\,B & 16\,483\,399 & 59\,140 & & 399 & 4 \\
 26 May & 18:26 & 03:19 & 8.7 & A,\,B & 452\,461 & 65\,551 & & 22 & 4 \\
 27 May & 19:09 & 02:50 & 7.5 & B & 2\,102\,839 & 60\,902 & & 69 & 3 \\
 \midrule
 20 Jun & 15:03 & 23:07 & 7.5 & A,\,B & 1\,014\,873 & 25\,962 & & 148 & 5 \\
 21 Jun & 15:35 & 00:10 & 8.3 & A,\,B & 723\,232 & 32\,432 & & 164 & 5 \\
 22 Jun & 17:21 & 00:12 & 6.3 & A,\,B & 1\,354\,033 & 13\,447 & & 68 & 7 \\
 23 Jun & 17:09 & 01:51 & 8.4 & B & 930\,107 & 35\,207 & & 55 & 4 \\
 \midrule
 17 Jul & 18:10 & 20:59 & 2.2 & B & 348\,323 & 14\,434 & & 11 & 2 \\
 18 Jul & 13:49 & 22:04 & 7.7 & A,\,B & 2\,370\,062 & 44\,735 & & 101 & 4 \\
 19 Jul & 15:01 & 23:05 & 7.7 & A,\,B & 20\,931\,824 & 57\,788 & & 60 & 2 \\
 20 Jul & 14:59 & 00:06 & 8.8 & A,\,B & 8\,812\,410 & 57\,052 & & 173 & 3 \\
 21 Jul & 16:02 & 01:02 & 8.9 & B & 13\,716\,161 & 62\,076 & & 59 & 2 \\
 \midrule
 14 Aug & 13:17 & 19:53 & 5.8 & A,\,B & 5\,302\,640 & 32\,248 & & 163 & 3 \\
 15 Aug & 12:39 & 20:57 & 7.2 & A,\,B & 197\,411\,\tnote{d}\hspace{-\tnotedlen} & 25\,872 & & 36 & 2 \\
 16 Aug & 13:02 & 21:57 & 8.7 & A,\,B & 4\,724\,087 & 36\,600 & & 92 & 6 \\
 \midrule
 10 Sep & 09:53 & 17:32 & 2.4 & A*,\,B* & 1\,437\,656 & 9\,994 & & 136 & 0 \\
 11 Sep & 10:11 & 18:41 & 8.3 & A*,\,B* & 9\,323\,492 & 38\,235 & & 798 & 3 \\
 12 Sep & 10:49 & 19:45 & 8.7 & A*,\,B* & 3\,206\,896 & 38\,709 & & 650 & 4 \\
 13 Sep & 11:43 & 20:48 & 4.0 & B* & 1\,094\,569 & 15\,178 & & 140 & 1 \\
 14 Sep & 12:32 & 21:46 & 8.4 & B* & 11\,571\,661 & 33\,108 & & 927 & 3 \\
 \midrule
 Total\,\tnote{b} & & & 148.7 & A,\,A*,\,B,\,B* & 123\,140\,611 & 794\,568 & & 4\,305 & 73 \\
 \bottomrule
\end{tabular}

  \begin{tablenotes}
   \titem{a} Due to time lost to configuration changes etc., the duration of active observations is typically slightly less than the time between the start and end of an observing run.
   \titem{b} Total does not include observations from April, which were excluded due to several technical problems (see text).
   \titem{c} Real-time anticoincidence logic was not fully active for these runs.
   \titem{d} Raw trigger rates were logged for only part of these observations, and have been extrapolated to fill the gaps.
  \end{tablenotes}
  \label{tab:obs}
 \end{threeparttable}
\end{table*}

Attenuation was set individually for each beam and polarisation to maintain an RMS signal power of \mbox{$\sim 10$}--11~ADU, despite variation in system temperature between on- and off-Moon beams.  This allowed peaks of up to \mbox{$\sim 12\sigma$} to be recorded without exceeding the 8-bit digitisation range of $-128$ to $+127$~ADU.  The attenuation settings were not adjusted to compensate for minor fluctuations in the instrumental gain, but were adjusted each time the pointing configuration was changed, moving beams between on- and off-Moon positions.

To permit the real-time anticoincidence filter to operate between the beams of the multibeam receiver, we calibrated their relative timing in each observing session using short-duration RFI pulses which were detected by all beams.  The precision of this calibration was determined by the timescale of the RFI pulses: typically 5--10~ns, which is small compared to the 200~ns exclusion window applied around each trigger by the anticoincidence filter.  The trigger threshold for the off-Moon beam was maintained at \mbox{$\sim 4.6\sigma$}, at which level the trigger rate was dominated by thermal noise, typically being \mbox{$\sim 10$}~kHz for both polarisations combined, although this varied significantly with small changes in the noise level.  This trigger rate leads to \mbox{$\sim 0.2$}\% of the observing time being excluded by the real-time anticoincidence filter (see \secref{sec:dutycycle}).  When a second off-Moon beam was employed (in configurations A* and B*; see \figref{fig:configs}), it was still treated as an on-Moon beam by the trigger logic (see \figref{fig:trigger_logic}), so it was set to a high threshold to prevent it from triggering and used for the exclusion of RFI only in retrospective processing (\secref{sec:cuts}).

\begin{figure}
 \centering
 \includegraphics[width=\linewidth]{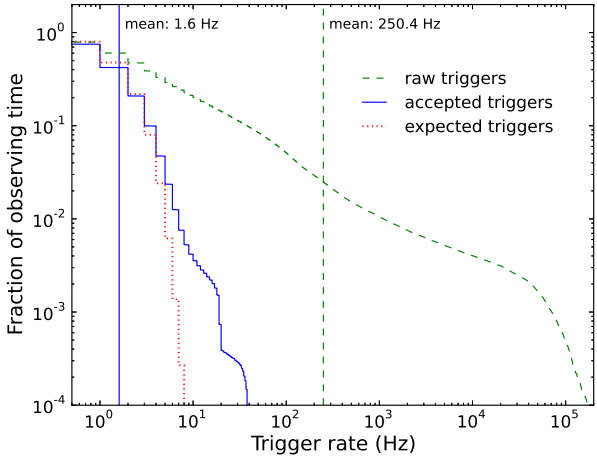}
 \caption{Cumulative histograms of the collective trigger rate of all on-Moon beams during our observations, both before the real-time anticoincidence filter (raw triggers; dashed) and after (accepted triggers; solid).  The expected distribution of trigger rates is also shown (dotted), based on Poisson statistics within each 1~s interval and with the same mean rate as the accepted triggers.  The raw trigger rate is typically elevated due to \gls{RFI}, but most of this excess is eliminated by the real-time anticoincidence filter, with the rate of accepted triggers conforming much more closely to the expected distribution.}
 \label{fig:trigrate}
\end{figure}

The threshold for triggers in the on-Moon beams, which cause events to be recorded, was maintained at \mbox{$\sim 6.4\sigma$}.  Pulses exceeding this threshold were typically dominated by impulsive RFI, with a highly variable raw trigger rate which averaged 251~Hz but occasionally exceeded 100~kHz, based on logs of the raw trigger rates during each second.  The majority of these raw triggers were excluded by the anticoincidence filter as shown in \figref{fig:trigrate}, and the events accepted by the filter were dominated by thermal noise, except for some short periods (total of several hours) during the September observing run with exceptionally high RFI activity, which were excluded at the time of the observations and not included in the analysis.  The remaining thermal noise events occurred at a collective rate of 1--2~Hz for all on-Moon beams, and were recorded for further analysis.

Apart from the signal optimisation described in \secref{sec:signal}, we applied additional processing to recorded events to correct for minor instrumental effects, described in \secref{sec:processing}.  High-significance events (\mbox{$\gtrsim 8\sigma$}) were dominated by RFI,
 so we applied various anti-RFI cuts to remove these, as described in \secref{sec:cuts}.  The remaining high-significance events occur at a low, constant rate which is consistent with the expected thermal noise.  The amplitude of the most significant event to pass all cuts defines the threshold sensitivity of this experiment, which we discuss in \secref{sec:threshold}.

\subsection{Processing}
\label{sec:processing}

Our data were affected by several minor bugs in the digital signal-processing firmware.  One occasionally caused a single sample value to be repeated, overwriting the remainder of a buffer; another caused pairs of adjacent samples to be swapped, rendering the output of the dedispersion filter incorrect.  Both of these affected only the initial observing run in April 2010, which was excluded from this analysis.  A persistent bug caused spurious values to appear at the start and end of each buffer, but the samples which meet the trigger condition occur in the centre of the buffer and are therefore unaffected, so we simply truncated the first and last eight samples in each case.

As the dedispersion filter is a novel system developed for this experiment, near the limits of the capabilities of the available digital signal-processing hardware, we tested the recorded data to ensure that it had performed as intended.  We were able to do this because each frame of recorded data, corresponding to a single event which passed the real-time anticoincidence filter, contains copies of the signal both before and after it was acted on by the filter, allowing us to apply a software reimplementation of the filter to the former copy to reproduce the latter.  Doing this, we found that the filter produced inconsistent results under certain circumstances.
\begin{itemize}
 \item In the April 2010 observing run, due to the sample-swapping bug described above, the filter acted on an out-of-order representation of the raw data, resulting in incorrect dedispersion.  This bug, however, already motivated the exclusion of this run from this analysis.
 \item In the March 2010 observing run, a small fraction (\mbox{$\sim 2$}\%) of recorded frames showed a minor mismatch between the output of the real-time dedispersion filter and its software implementation acting on the raw data, with occasional samples differing by 1--2~ADU (\mbox{$\sim 0.1$}--$0.2\sigma$).  The cause of this bug is unknown, but we consider it insignificant and ignore it.
 \item On each occasion when the dedispersion filter setting was changed, a single frame was not processed correctly under either the old or the new setting.  These occasions were extremely infrequent, and excluded from further analysis.
 \item As both the raw and filtered data were recorded as 8-bit integers, the filter was unable to function correctly when dedispersion would increase the amplitude of a peak beyond the range $-128$ to $+127$~ADU (\mbox{$\pm \sim 12\sigma$}).  These events were still recorded, and the raw data is still valid.
\end{itemize}
Partly because of this last issue, further analysis was performed only on the raw (rather than dedispersed) recorded data.  This also simplified bandpass optimisation, which was performed according to the instrumental bandpass rather than the bandpass after the action of the real-time dedispersion filter; and dedispersion, which was based directly on the retrospective GPS-derived STEC measurements, rather than the difference between these and the varying real-time dedispersion filter setting.

Analog-to-digital converters (ADCs) do not typically display perfectly linear behaviour.  The non-linearity of the ADCs used in this experiment is known to cause errors of up to 2.5\% in the recorded pulse amplitudes\gcitep{bray2012}.  This is much less significant than the non-linearity observed earlier in the signal path (see \secref{sec:linearity}), so it is not necessary to correct for it in real time.  However, a small error in the pulse amplitude can lead to a large error in the number of recorded high-significance events, so it is desirable to remove the effect in retrospective processing.  Since the magnitude of the effect is different in each ADC, it causes differences in the number of high-significance events recorded in each beam, which we initially mistook for a real phenomenon rather than an instrumental effect.  We calibrated the errors individually for each ADC and used these measurements to perform the linearity correction by rescaling the pulse amplitudes.  For simplicity, we applied a correction averaged between positive and negative directions in voltage, neglecting the minor (\mbox{$\pm 0.4$}\%) variation between them.

Spectra from recorded events show common features from instrumental effects: at zero frequency, resulting from a direct-current offset in the ADCs; and at the maximum IF frequency (512~MHz), which is a harmonic of the operating frequency of the digital electronics.  In real time, these were strongly suppressed by the dedispersion filter.  For retrospective analysis, they were completely removed.

Although RFI is of greatest concern in this experiment when it is impulsive in nature and may be mistaken for an Askaryan pulse, continuous narrow-band RFI is also an unwanted source of noise.  We dealt with this simply, removing any channels in the Fourier transforms of the data with a spectral density exceeding four times the RMS value.  These constitute a smaller fraction of the band than in lower-frequency observations \citep[e.g.][]{buitink2010} --- typically 0--2 out of 2,041 channels --- and their removal does not appreciably affect the sensitivity.

After the above steps and the bandpass optimisation from \secref{sec:bandpass}, the RMS signal power for each beam and polarisation was calculated and averaged over each 1-minute interval of observing time, to avoid the \mbox{$\sim 2$}\% uncertainty associated with calculating this quantity from a single buffer of data while still measuring fluctuations in the instrumental gain over longer timescales.  We then performed the remaining optimisation steps from \secref{sec:signal}, including dedispersion based on STEC values calculated once per minute of observing time.  The magnitude of the most extreme peak in the data after this last step, divided by the RMS signal power, gives us the significance of the peak $n_\sigma$, as defined in \eqnref{eqn:nsigma}, in units of $\sigma$.

\subsection{Removal of RFI}
\label{sec:cuts}

Some RFI events escaped the real-time anticoincidence filter: although they featured peaks in multiple beams, the relative amplitudes of these peaks varied, so the triggering peak could exceed the trigger threshold in one beam while the corresponding peaks in the other beams remained below the threshold.  These constituted a small fraction of the raw triggers, but were still numerous enough to dominate the population of recorded high-significance events, limiting the sensitivity of the experiment.  To exclude them, we applied a series of cuts, tightening the exclusion criteria used in the real-time filter and discriminating against other features observed in the RFI events.  As we have no expectation for the spectrum or dispersion of an RFI pulse, these cuts were applied to the raw data with no dedispersion or bandpass optimisation; but they did include interpolation, formation of the signal envelope, and other corrections that were not performed in real time.

The initial \textsl{anticoincidence} cuts are a simple refinement of the real-time filter, exploiting the length of the buffers of recorded data (2--4~$\mu$s) and including the processing described in \secref{sec:processing}.  They excluded
\begin{cutenumerate}
 \item events with a peak in excess of 6.2$\sigma$ in any beam other than the triggering beam; and \label{it:cut_2beam}
 \item events with peaks in excess of 5.8$\sigma$ in any two beams, or both polarisations of the same beam, other than the triggering beam, \label{it:cut_3beam}
\end{cutenumerate}
which together constituted the majority of the recorded RFI.  These thresholds were chosen as a compromise between discriminating power and the false exclusion rate: the probability that a genuine event will be misidentified as RFI and thus excluded.  The multi-beam requirement in \cutref{it:cut_3beam} allows the threshold to be set lower than in \cutref{it:cut_2beam}.

Visual inspection of RFI events, including those which passed \cutsref{it:cut_2beam}{it:cut_3beam}, frequently showed extended structure in the time domain: both broad pulses with a duration of several tens of ns and series of pulses with regular spacings of several hundreds of ns.  We were able to experimentally reproduce the characteristic latter profile with the radio emission from an electric barbecue lighter, and we ascribe these events to other anthropogenic electrical activity with similar qualities.  This motivated further cuts on the \textsl{width} of the pulses in a single beam in a single event, excluding
\begin{cutenumerate}
 \item events with a peak in excess of 6.2$\sigma$ in the same beam and polarisation as the peak which met the trigger condition, but separated from it by more than 10~ns; and \label{it:cut_width}
 \item events with a peak in excess of 6.2$\sigma$ in the same beam as the peak which met the trigger condition, but in the other polarisation, and separated from it by more than 60~ns. \label{it:cut_polwidth}
\end{cutenumerate}
These cuts discriminated both against repeated pulses and, to a lesser extent, broad single pulses.  \Cutref{it:cut_width} places an effective 10--20~ns upper limit on the width of a pulse for it to avoid being excluded as RFI, but this is consistent with the expected \mbox{$\sim 1$}~ns duration of an Askaryan pulse.  The increased separation permitted in \cutref{it:cut_polwidth} is to allow for the remote possibility of a large error in the timing calibration between polarisations.

The cuts applied thus far have relatively lax thresholds and hence a low false exclusion rate, which we quantify with two techniques.  The first is to determine the number of samples to which the exclusion threshold has been applied in each cut, and to determine the probability that it has been exceeded based on their expected Gaussian distribution and the consequent Rayleigh distribution of the signal envelope.  The second is to operate our instrument with a minimal trigger threshold, continuously recording noise data without the usual bias towards high-amplitude peaks, and subjecting this noise data to the same processing as the observational data.  Applying the cuts to the noise data and measuring the fraction of excluded events gives a measure of the false exclusion rate which is sensitive to non-Gaussianness in the background noise, from both RFI and instrumental effects.  Both approaches yield for \cutsref{it:cut_2beam}{it:cut_polwidth} a collective false exclusion rate of $<0.1$\%, leading to a correspondingly insignificant loss of observing time.

\begin{figure}
 \centering
 \includegraphics[width=\linewidth]{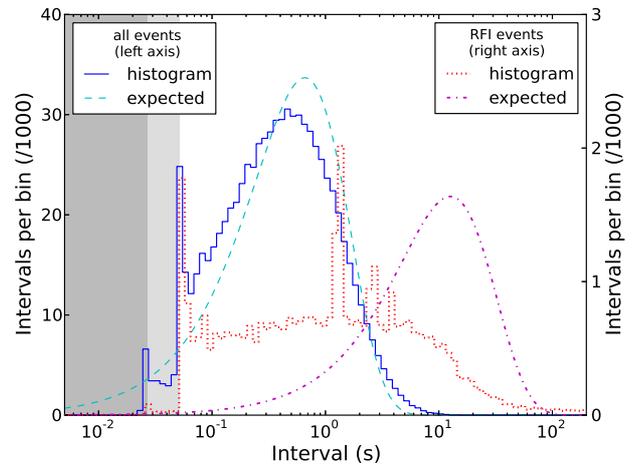}
 \caption{Distribution of intervals between events, for all recorded events (solid; left axis) and for events identified as RFI by \cutsref{it:cut_2beam}{it:cut_polwidth} (dotted; right axis) with a bin width of 0.05~decades.  These two histograms have been slightly offset for display purposes.  The expected distributions for a purely stochastic distribution of events are also shown.  The distribution for all recorded events, which are dominated by thermal noise, fits quite closely to the expected distribution, with minor deviations caused by variation in the trigger rate during our observations.  The events identified as RFI are typically separated by shorter intervals than expected, indicating that they are clustered in time.  The shaded regions are excluded due to the maximum possible trigger rates for buffers of length 2~$\mu$s (dark) and 4~$\mu$s (light); the pile-up at their boundaries results from short periods when RFI or a decreased trigger threshold caused our instrument to trigger continuously.  Peaks in the distribution for RFI events are visible at a period of 1.3~s and its subharmonics, indicating a source of RFI pulses with regular spacing of this width.}
 \label{fig:timegaps}
\end{figure}

\begin{figure}
 \centering
 \includegraphics[width=\linewidth]{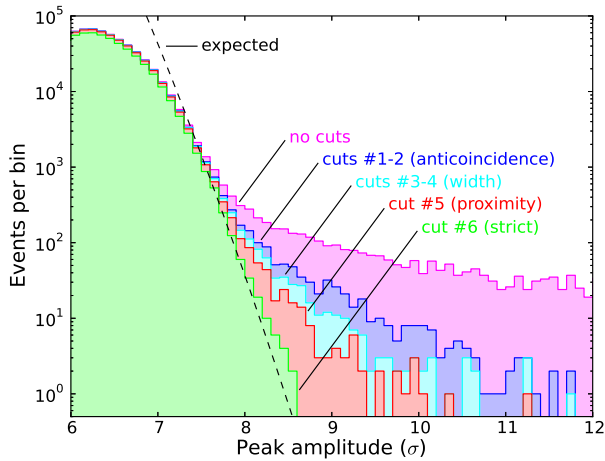}
 \caption{Distribution of amplitudes of events (bin width 0.1$\sigma$) for observing runs from May 2010 onwards, after each set of anti-RFI cuts except for \cutref{it:cut_spect}, which has only a minimal effect (see \tabref{tab:cuts}).  The expected distribution incorporates the effects of interpolation and formation of the signal envelope; the smearing effect due to random error from quantisation noise and variation in the RMS signal power \citep[as per][]{bray2012} and from the differential non-linearity between positive and negative voltage neglected in the earlier correction for ADC non-linearity; and the reduction in effective observing time described in \secref{sec:dutycycle}, including the effects of the false exclusion rates for \cutsref{it:cut_2beam}{it:cut_spect}.  The deficit in the number of events at low amplitudes is due to these events not being consistently recorded in this experiment.  The number of events with low amplitudes, which are dominated by thermal noise, is noticeably decreased by \cutref{it:cut_strict}, due to its high false exclusion rate.  After all cuts, the number of events at high amplitudes is consistent with expectations.}
 \label{fig:hist_byfilt}
\end{figure}

The events excluded as RFI by \cutsref{it:cut_2beam}{it:cut_polwidth} are clustered in time on scales \mbox{$\lesssim 10$}~s (see \figref{fig:timegaps}), indicating that they originate from transient sources which are typically active for a similar period.  The events which pass these cuts are significantly in excess of those expected from pure noise for peak amplitudes \mbox{$\gtrsim 8\sigma$} (see \figref{fig:hist_byfilt}), suggesting that they are dominated by RFI which escaped the cuts by not displaying the characteristics typical of RFI events: for example, having an origin in a sidelobe of one beam, but in nulls of the sidelobe patterns of the other beams, so that they do not detect coincident pulses.  Based on this, we expect RFI events which passed the cuts to also display clustering behaviour with one another; but, as the positions of the sidelobes change over time (relative to local RFI sources, as the telescope tracks the Moon), they will generally occur at different times from RFI events which were already excluded, and not be clustered with them.  This motivates a further cut based on the \textsl{proximity} of events to one another in time, in which we exclude
\begin{cutenumerate}
 \item events recorded within 10~s of another event with a peak exceeding 8$\sigma$, which also passed \cutsref{it:cut_2beam}{it:cut_polwidth}. \label{it:cut_prox}
\end{cutenumerate}
We choose to apply an exclusion window only around events with peak amplitude \mbox{$> 8\sigma$} because we expect these events to be dominated by RFI; and we ignore events excluded by previous cuts because, as argued above, we do not expect these to be clustered with RFI events remaining in our sample.  In this way, we limit ourselves to exclusion windows of length 20~s centred on each of 727 events.  These are, as expected, clustered, so the windows overlap and the total excluded time is only $1.1 \times 10^4$~s, equivalent to a 1.7\% false exclusion rate.

The clustering exploited in \cutref{it:cut_prox} is too weak to act as a completely reliable discriminant against RFI, so we perform an additional cut to remove events with very weak coincident pulses in multiple beams.  To minimise the false exclusion rate associated with a lower exclusion threshold, this \textsl{strict} cut searches only for very closely coincident pulses, excluding
\begin{cutenumerate}
 \item events with a 4.5$\sigma$ peak within 20~ns of the position of the peak in the triggering beam, in another beam. \label{it:cut_strict}
\end{cutenumerate}
This \mbox{$\pm 20$}~ns window is sufficiently wide to encompass the likely timing calibration error between beams, while minimising the probability that it will contain a peak exceeding the exclusion threshold purely from Gaussian noise.  Nonetheless, due to the low threshold, this cut has the highest false exclusion rate: calculating this as for \cutsref{it:cut_2beam}{it:cut_polwidth}, and conservatively taking the higher value produced by applying the cut to experimental noise data, the false exclusion rate is 5.4\%.  This cut also limits the sensitivity of this experiment due to the possibility of a lunar Askaryan pulse being seen in multiple beams and excluded as RFI.

\Cutsref{it:cut_2beam}{it:cut_strict} were sufficient to exclude impulsive RFI, but a further cut was necessary to identify periods of continuous narrow-band RFI which exceeded the dynamic range.  A fluctuation in the voltage beyond the limits of the 8-bit digitisation range will be clipped to $+127$~ADU (if positive) or $-128$~ADU (if negative), and its full magnitude will be unrecorded.  Powerful narrow-band RFI causes this clipping to happen frequently; although it is removed in retrospective processing, the problem occurs earlier, at the digitisation stage, so the information is already lost.  We are therefore unable to reliably record high-significance events during periods of intense narrow-band RFI, so we remove these periods with a further cut on the \textsl{spectrum} of pulses, excluding
\begin{cutenumerate}
 \item events with more than 50\% of the spectral power concentrated in less than 1\% of the band, in any beam or polarisation. \label{it:cut_spect}
\end{cutenumerate}
Unlike previous cuts, this excludes very few high-significance events, but it does exclude all of the 24 events passing \cutsref{it:cut_2beam}{it:cut_strict} with sample values that reach the limits of the digitisation range, which may have been clipped.  This ensures that any remaining high-significance events are recognised as such, rather than potentially being reconstructed with lower significance because their peak amplitude was reduced by clipping.  The false exclusion rate, calculated as for previous cuts, is 0.7\%.

\begin{table}
 \centering
 \begin{threeparttable}
  \caption{Numbers of events remaining after each set of anti-RFI cuts.}
  \begin{tabular}{ccrr}
 \toprule
 \multirow{2}{*}{Cut(s)} & \multirow{2}{*}{Criterion} & \multicolumn{2}{c}{Events} \\
 \cmidrule{3-4}
  & & \multicolumn{1}{c}{total} & \multicolumn{1}{c}{$> 8\sigma$} \\
 \midrule
 --- & none & 794\,568 & 4\,305 \\
 \#1--2 & anticoincidence & 756\,947 & 789 \\
 \#3--4 & width & 755\,867 & 483 \\
 \#5 & proximity & 741\,045 & 294 \\
 \#6 & strict & 668\,090 & 74 \\
 \#7 & spectrum & 657\,421 & 73 \\
 \bottomrule
\end{tabular}

  \label{tab:cuts}
 \end{threeparttable}
\end{table}

\Figref{fig:hist_byfilt} shows the distribution of the peak amplitudes after complete processing, including dedispersion and bandpass optimisation, and the effects on this distribution of the series of cuts described above.  The numbers of events after each cut are also listed in \tabref{tab:cuts}.  Events with amplitudes \mbox{$> 8\sigma$} are initially dominated by RFI, but after all cuts there are only 73 such events against an expectation of \mbox{$61 \pm 8$}.  Of these, 3 events appear to have characteristics --- broad, coincident or repeated pulses --- suggesting that they may also be due to RFI, but the significance of these features cannot be rigorously established through visual inspection, so we do not exclude these events.  The remaining 70 events all appear to be consistent with the expected rare fluctuations in the background thermal noise.

The most significant event remaining after cuts has an amplitude of $8.6\sigma$, with 3,000 more significant events being excluded.  The real-time anticoincidence filter excluded raw triggers at a ratio of \mbox{$\sim 150:1$} (see \tabref{tab:obs}).  As the amplitudes of events excluded in real time are not recorded, we cannot state with certainty the total number of excluded events with amplitudes exceeding the 8.6$\sigma$ significance threshold; but, if the distribution of amplitudes is the same for raw triggers as for recorded events, then the combination of the real-time filter and subsequent cuts has excluded \mbox{$\sim 450,000$} high-significance RFI events and accepted none.  Since raw triggers are dominated by RFI, we expect them to be biased towards larger amplitudes, in which case the number of successfully rejected high-significance events is even greater than this.

\subsection{Threshold}
\label{sec:threshold}

The 8.6$\sigma$ peak amplitude of the most significant event remaining after cuts defines the significance threshold of this experiment.  If an event had been detected in excess of this threshold, the confidence of its identification as a lunar-origin Askaryan pulse would not be rigorously established, as the cuts were not all defined \textsl{a priori}.  However, the absence of such events allows us to place a limit on the intensity of radio pulses originating from the Moon during this experiment; which, combined with the rigorous testing of the false exclusion rates of the cuts to determine the effective observing time, allows limits to be placed on the fluxes of UHE cosmic rays and neutrinos.

Thus far, however, we have only established that there are no events exceeding this threshold in the recorded data.  We must also establish that any such event occurring during this experiment would have been recorded in the first place.  Neglecting effects which merely reduce the effective observing time, which we discuss in \secref{sec:dutycycle}, there are two possible cases in which we could fail to record a high-significance event.

In the first case, a pulse might be properly sampled and processed, but its peak amplitude imperfectly reconstructed by the suboptimal real-time processing, so that it falls below the trigger threshold and is not recorded.  The maximum loss of amplitude in real-time reconstruction relative to full retrospective processing is 23\%, as found in \secref{sec:signal_sens}.  For a peak with a full magnitude of 8.6$\sigma$ or greater, this means that it will be detected in real time with a minimum magnitude of 6.6$\sigma$.  Since this is still above the 6.4$\sigma$ trigger threshold, all such events are recorded and will be correctly analysed retrospectively, so this case does not occur.

In the second case, a pulse might exceed the \mbox{$\pm 12\sigma$} digitisation range of the ADCs, causing it to be clipped and its full magnitude to go unrecorded, with its amplitude then further decreased by the real-time dedispersion filter to a level below the trigger threshold.  While it is not generally expected for dedispersion to reduce the amplitude of a dispersed lunar-origin Askaryan pulse as required in this case, note that both clipping and the surrounding thermal noise may alter the inherent dispersion of the pulse, and the real-time dedispersion filter may change both the phase of the pulse and the phase of the sampling times relative to its peak, with consequent effects on its amplitude.  Taking this reduction in amplitude, for the maximum setting of the filter, to be equal to the expected loss from dispersion for the maximum ionospheric STEC during our observations, and including the effects of partial interpolation, the maximum signal loss is 32\%, found with the same simulation used in \secref{sec:signal_sens}.  This is sufficient to reduce the amplitude of a clipped pulse from 12$\sigma$ to 8.2$\sigma$.  As this is still above the 6.4$\sigma$ trigger threshold, all such events are recorded, and this case also does not occur; so the absence of any clipped pulses after \cutref{it:cut_spect} implies that clipped events occurred only as RFI.

With these possibilities eliminated, we have established that no lunar-origin pulses occurred during our experiment (within the effective observing time; see \secref{sec:dutycycle}) with a reconstructed pulse height in excess of 8.6$\sigma$.  This threshold does not allow for loss of amplitude due to suboptimality of the final reconstruction (\mbox{$< 0.4$}\%; see \secref{sec:signal_sens}), or from Faraday rotation (\mbox{$\lesssim 1$}\%; see \secref{sec:pointing}), but these are small compared to the calibration uncertainty of $\pm 4.5\%$ (random) $\pm 4\%$ (systematic) found for the electric field in \secref{sec:calibration}.  We convert this threshold significance \mbox{$n_\sigma = 8.6$} (see \eqnref{eqn:nsigma}) into a threshold spectral electric field strength with \eqnref{eqn:thresh}, using the calibrated sensitivity for each beam pointing and polarisation.

For a limb beam pointing (see \figref{fig:configs}), this gives a threshold of 0.0047 $\mu$V/m/MHz for a pulse originating at the centre of the beam, polarised radially to the Moon.  At the closest point on the lunar limb, the threshold is 0.0053 $\mu$V/m/MHz for the same polarisation.  Further away on the limb, the sensitivity to a radially-polarised pulse decreases, both due to the greater distance from the centre of the beam and due to misalignment between the polarisation of the beam and that of the pulse.  The fraction of the lunar limb over which the threshold takes no more than $\sqrt{2}$ times its minimum value on the limb (i.e.\ at least half sensitivity in the power domain) is 10\% for a single limb beam.

\begin{figure}
 \centering
 \includegraphics[width=\linewidth]{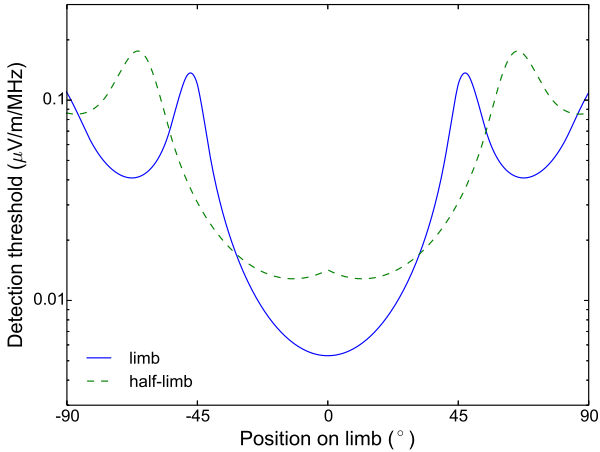}
 \caption{Detection threshold for a radially-aligned pulse along the lunar limb for the limb and half-limb beams (see \figref{fig:configs}).  A limb position of 0\degree\ corresponds to a pulse originating at the point on the limb closest to the beam.  The limb beam is most sensitive, but the sensitivity declines rapidly for a pulse away from this point, which is both more distant from the centre of the beam and suboptimally aligned for the radial polarisation channel.  The half-limb beam is less sensitive, but a pulse slightly removed from the closest point on the limb will be better aligned for one or the other polarisation channel, resulting in greater limb coverage.}
 \label{fig:thresholds}
\end{figure}

For a half-limb beam pointing, the threshold is 0.0074 $\mu$V/m/MHz for either polarisation (both at 45\degree\ to a radial alignment) at the centre of the beam, or 0.010 $\mu$V/m/MHz at the closest point on the limb.  This is equivalent to a threshold of 0.014 $\mu$V/m/MHz for a radial pulse at this point.  Further away on the limb, a radially aligned pulse more closely matches one or the other polarisation of the beam, which results in greater limb coverage: 20\%, calculated as for the limb beam.  These limb coverage values are suitable for inclusion in particle aperture models such as that of \citet{gayley2009}.  The limb coverage for both beams is shown in \figref{fig:thresholds}.

For comparison, the two previous lunar radio experiments with the lowest detection thresholds are LUNASKA ATCA with a threshold of 0.0145--0.016 $\mu$V/m/MHz\gcitep{james2010} and RESUN with a threshold of 0.017 $\mu$V/m/MHz\gcitep{jaeger2010}.  (\citet{jaeger2010} report a threshold of 0.013 $\mu$V/m/MHz for the Kalyazin experiment\gcitep{beresnyak2005}, but this is not compatible with the originally published flux density threshold of 13.5~kJy.)  Consequently, the threshold for the radial polarisation of our limb beams improves over previous experiments by a factor of three.  This results in a decrease by a corresponding factor in the minimum detectable particle energy compared to these previous experiments.

Note that each of these thresholds represents the spectral electric field strength for which a pulse has a 50\% probability of detection: the addition of thermal noise may cause pulses over the threshold to be missed, or pulses under the threshold to be detected.  The amplitude of the noise is \mbox{$1/8.6 \sim 12\%$} of the threshold value.  The effect of thermal noise on the aperture for UHE particles is incorporated into simulations such as those of \citet{james2009b}, although \citet{gayley2009} argue that the effect is negligible.

\subsubsection{Effective observing time}
\label{sec:dutycycle}

The effective observing time is decreased from the total time reported in \tabref{tab:obs}, either through a decrease in the duty cycle or a chance of incorrectly excluding a detected Askaryan pulse, by three effects.
\begin{itemize}
 \item The real-time anticoincidence filter excludes events in which the trigger threshold is exceeded in multiple beams, as described in \secref{sec:design}.  It applies a 200~ns exclusion window around each trigger, with the trigger rate being dominated by the off-Moon beam.  The number of such triggers is logged, so an upper limit (assuming no overlapping windows) to the total excluded time is known.
 \item After a successful trigger, the system is temporarily unable to respond to a further trigger as it copies the previous event to permanent storage, as described in \secref{sec:design}.  We measured the length of this interval by operating briefly with a minimal trigger threshold to cause continuous triggering, and noting the number of recorded events.  The inverse of this event rate gives us the excluded interval after each recorded event, which is 27~ms when the buffer length is set to 2~$\mu$s and 51~ms for a buffer length of 4~$\mu$s.
 \item Each anti-RFI cut has a corresponding false exclusion rate as described in \secref{sec:cuts}, evaluated primarily by applying the same cuts to a sample of background noise data.
\end{itemize}
We consider the above effects separately for observing time with two on-Moon beams (the September observing run) and three on-Moon beams (all other observing runs), with the results shown in \tabref{tab:dur}.  RFI was more prevalent in the September observing run, but was more effectively filtered out by our initial anticoincidence and width cuts, probably due to the use of an additional off-Moon beam, so less time was lost to the proximity cut than in other observing runs.  The total effective observing time, combining all observing runs, is 127.2 hours with each of two limb beams and 99.4 hours with one half-limb beam.

\begin{table}
 \centering
 \begin{threeparttable}
  \caption{Effective duration of observations, after losses due to the real-time anticoincidence filter, time taken for data storage, and false exclusion rates of anti-RFI cuts.}
  \begin{tabular}{lrr}
 \toprule
 Configurations & A,\,B\,\tnote{a}\hspace{-\tnotealen} & A*,\,B*\,\tnote{b}\hspace{-\tnoteblen} \\
 \midrule
 Total duration & 116.9 hrs & 31.8 hrs \\
 \midrule
 Real-time anticoincidence & $-$0.4\% & $-$0.2\% \\
 Data storage & $-$6.8\% & $-$6.1\% \\
 Anti-\gls{RFI} cuts & $-$8.4\% & $-$6.7\% \\
 \midrule
 Effective duration & 99.4 hrs & 27.8 hrs \\
 \bottomrule
\end{tabular}

  \label{tab:dur}
  \begin{tablenotes}
   \titem{a} With two limb beams and one half-limb beam.
   \titem{b} With two limb beams.
  \end{tablenotes}
 \end{threeparttable}
\end{table}

\section{Conclusion and summary}
\label{sec:conclusion}

We have conducted an experiment to search for radio pulses from particle cascades initiated by UHE particles interacting in the Moon.  Our anticoincidence filter and subsequent cuts exclude effectively all RFI, and the remaining events are consistent with those expected from thermal noise.  We detected no pulses originating on the limb of the Moon with an electric field exceeding 0.0053 $\mu$V/m/MHz, assuming linear polarisation radial to the Moon, during 127 hours of effective observing time.  This threshold is an improvement over previous lunar radio experiments by a factor of three.  Our non-detection implies limits on the fluxes of UHE cosmic rays and neutrinos, which may be modeled based on the parameters reported here.

We have achieved, for the first time, effectively complete exclusion of RFI pulses without the benefit of a coincidence requirement between multiple antennas or bands.  Previous experiments have either required a coincidence between multiple channels, preventing them from achieving the full coherent sensitivity possible with their combined bandwidth and collecting area, or been dominated by remnant RFI.

\subsection{Considerations for future experiments}

Future lunar radio experiments will take advantage of new and improved radio telescopes which are currently under development, achieving greater sensitivity to UHE particles.  In particular, \citet{singh2012} have proposed lunar radio observations with LOFAR, an aperture array telescope currently in its commissioning phase, and \citet{bray2013} have proposed continued use of the Parkes radio telescope used in this experiment, using one of the Phased Array Feed (PAF) receivers recently developed for ASKAP\gcitep{schinckel2012}.  Both of these proposed experiments would utilise multiple beams pointing on the Moon with a real-time anticoincidence filtering scheme to identify lunar-origin Askaryan pulses.  Here we summarise the experimental considerations addressed in this work that may be relevant for these and other future experiments.

The optimum signal path for a lunar radio experiment can be understood in terms of the aspects of a matched filter, with the phase, dispersion and bandpass optimised to match the expected characteristics of an Askaryan pulse.  In each of these respects, this experiment improves on previous work.
\begin{description}
 \item[Phase:] In typical radio receivers, the signal is shifted to an intermediate frequency by mixing it with a local oscillator signal, effectively randomising the phase of a pulse and potentially decreasing its peak amplitude (by up to 18\%, in this case).  We form the signal envelope to restore the maximum peak amplitude, but this also increases the noise level, partly offsetting the improvement in the signal-to-noise ratio.  In some recent radio telescopes the radio frequency signal is sampled directly; provided that it is Nyquist-sampled (i.e.\ no aliasing occurs), this preserves the original phase of a pulse.  The inherent phase of an Askaryan pulse corresponds to the worst-case peak amplitude; but, in this case, the maximum peak amplitude can be restored by performing a Hilbert transform on the signal, with no change in the noise level.
 \item[Dispersion:] Our experiment is the first to use a dedispersion filter which was adjusted in real time based on changes in the ionospheric STEC along the line of sight to the Moon.  The precision of this correction is limited by the availability of precise STEC measurements, particularly in real time.  The most precise values available are from GPS data, with the uncertainty dominated by the conversion from STEC (between ground station and GPS satellite) to VTEC and back.  Future experiments, which will generally be more affected by dispersion due to increased bandwidth, could directly use GPS STEC values measured from a ground station located at the telescope site, or measure the STEC through Faraday rotation of polarised lunar thermal emission as proposed by \citet{mcfadden2011}.
 \item[Bandpass:] For this experiment, we optimised the bandpass for sensitivity to lower-energy particles.  The improvement in sensitivity in this case is an insignificant 1--2\%, but will become increasingly significant for future experiments with larger fractional bandwidths.
\end{description}

The loss of sensitivity due to a finite sampling rate can be negated by interpolating to reconstruct intermediate sample values, as first suggested in this context by \citet{james2010}.  For a triggered experiment, this interpolation may need to be at least partially implemented in real time, as in this work.  The potential loss of sensitivity without this correction depends on the oversampling factor, and is 22\% for our experiment.

An improvement in sensitivity is obtained by aligning a linearly-polarised receiver with the expected polarisation of an Askaryan pulse, at least when observing at high radio frequencies for which Faraday rotation is minimal.  For future experiments, with sufficiently precise calibration of the delay and phase between polarisation channels of each beam, it may be possible to construct signals corresponding to arbitrary linear polarisations, allowing this benefit to be obtained regardless of the orientation and native polarisation of the receiver.

An understanding of the above considerations is required in order to properly evaluate the sensitivity of a lunar radio experiment to Askaryan pulses.  We have described techniques which allow this sensitivity to be maximised.  The implementation of these techniques in future lunar radio experiments, and eventually with the SKA, will result in the greatest possible aperture for the detection of UHE particles.

\section*{Acknowledgements}

The Parkes radio telescope is part of the Australia Telescope which is funded by the Commonwealth of Australia for operation as a National Facility managed by CSIRO.  Some data used in this study were acquired as part of NASA's Earth Science Data Systems and archived and distributed by the Crustal Dynamics Data Information System (CDDIS).  This research was supported by the Australian Research Council's Discovery Project funding scheme (project number DP0881006).  JDB acknowledges support from ERC-StG 307215 (LODESTONE).

\bibliographystyle{elsarticle-num-names}
\bibliography{all}

\begin{thebibliography}{43}
\providecommand{\natexlab}[1]{#1}
\providecommand{\url}[1]{\texttt{#1}}
\providecommand{\urlprefix}{URL }
\expandafter\ifx\csname urlstyle\endcsname\relax
  \providecommand{\doi}[1]{doi:\discretionary{}{}{}#1}\else
  \providecommand{\doi}[1]{doi:\discretionary{}{}{}\begingroup
  \urlstyle{rm}\url{#1}\endgroup}\fi
\providecommand{\bibinfo}[2]{#2}

\bibitem[{{Pierre Auger Collaboration}(2004)}]{abraham2004}
\bibinfo{author}{{Pierre Auger Collaboration}}, \bibinfo{title}{{Properties and
  performance of the prototype instrument for the Pierre Auger Observatory}},
  \bibinfo{journal}{\nima{}} \bibinfo{volume}{523} (\bibinfo{year}{2004})
  \bibinfo{pages}{50--95}.

\bibitem[{{Telescope Array Collaboration}(2012)}]{abu-zayyad2012c}
\bibinfo{author}{{Telescope Array Collaboration}}, \bibinfo{title}{{The surface
  detector array of the Telescope Array experiment}},
  \bibinfo{journal}{\nima{}} \bibinfo{volume}{689} (\bibinfo{year}{2012})
  \bibinfo{pages}{87--97}.

\bibitem[{{Dagkesamanskii} and {Zheleznykh}(1989)}]{dagkesamanskii1989}
\bibinfo{author}{R.~D. {Dagkesamanskii}}, \bibinfo{author}{I.~M. {Zheleznykh}},
  \bibinfo{title}{{Radio-astronomy method for detecting neutrinos and other
  elementary particles of superhigh energy}}, \bibinfo{journal}{\spjetpl{}}
  \bibinfo{volume}{50} (\bibinfo{year}{1989}) \bibinfo{pages}{259--261}.

\bibitem[{{Askaryan}(1962)}]{askaryan1962}
\bibinfo{author}{G.~A. {Askaryan}}, \bibinfo{title}{{Excess Negative Charge of
  an Electron-Photon Shower and its Coherent Radio Emission}},
  \bibinfo{journal}{\spjetp{}} \bibinfo{volume}{14}~(\bibinfo{number}{2})
  (\bibinfo{year}{1962}) \bibinfo{pages}{441--443}.

\bibitem[{{James} and {Protheroe}(2009{\natexlab{a}})}]{james2009b}
\bibinfo{author}{C.~W. {James}}, \bibinfo{author}{R.~J. {Protheroe}},
  \bibinfo{title}{{The sensitivity of the next generation of lunar Cherenkov
  observations to UHE neutrinos and cosmic rays}}, \bibinfo{journal}{\app{}}
  \bibinfo{volume}{30} (\bibinfo{year}{2009}{\natexlab{a}})
  \bibinfo{pages}{318--332}.

\bibitem[{{ter Veen} et~al.(2010){ter Veen}, {Buitink}, {Falcke}, {James},
  {Mevius}, {Scholten}, {Singh}, {Stappers}, and {de Vries}}]{terveen2010}
\bibinfo{author}{S.~{ter Veen}}, \bibinfo{author}{S.~{Buitink}},
  \bibinfo{author}{H.~{Falcke}}, \bibinfo{author}{C.~W. {James}},
  \bibinfo{author}{M.~{Mevius}}, \bibinfo{author}{O.~{Scholten}},
  \bibinfo{author}{K.~{Singh}}, \bibinfo{author}{B.~{Stappers}},
  \bibinfo{author}{K.~D. {de Vries}}, \bibinfo{title}{{Limit on the
  ultrahigh-energy cosmic-ray flux with the Westerbork synthesis radio
  telescope}}, \bibinfo{journal}{\prd{}}
  \bibinfo{volume}{82}~(\bibinfo{number}{10}) (\bibinfo{year}{2010})
  \bibinfo{pages}{103014}.

\bibitem[{{Jeong} et~al.(2012){Jeong}, {Reno}, and {Sarcevic}}]{jeong2012}
\bibinfo{author}{Y.~S. {Jeong}}, \bibinfo{author}{M.~H. {Reno}},
  \bibinfo{author}{I.~{Sarcevic}}, \bibinfo{title}{{Radio Cherenkov signals
  from the Moon: Neutrinos and cosmic rays}}, \bibinfo{journal}{\app{}}
  \bibinfo{volume}{35} (\bibinfo{year}{2012}) \bibinfo{pages}{383--395}.

\bibitem[{{Hankins} et~al.(1996){Hankins}, {Ekers}, and
  {O'Sullivan}}]{hankins1996}
\bibinfo{author}{T.~H. {Hankins}}, \bibinfo{author}{R.~D. {Ekers}},
  \bibinfo{author}{J.~D. {O'Sullivan}}, \bibinfo{title}{{A search for lunar
  radio \u{C}erenkov emission from high-energy neutrinos}},
  \bibinfo{journal}{\mnras{}} \bibinfo{volume}{283} (\bibinfo{year}{1996})
  \bibinfo{pages}{1027--1030}.

\bibitem[{{Gorham} et~al.(2004){Gorham}, {Hebert}, {Liewer}, {Naudet},
  {Saltzberg}, and {Williams}}]{gorham2004a}
\bibinfo{author}{P.~W. {Gorham}}, \bibinfo{author}{C.~L. {Hebert}},
  \bibinfo{author}{K.~M. {Liewer}}, \bibinfo{author}{C.~J. {Naudet}},
  \bibinfo{author}{D.~{Saltzberg}}, \bibinfo{author}{D.~{Williams}},
  \bibinfo{title}{{Experimental Limit on the Cosmic Diffuse Ultrahigh Energy
  Neutrino Flux}}, \bibinfo{journal}{\prl{}}
  \bibinfo{volume}{93}~(\bibinfo{number}{4}) (\bibinfo{year}{2004})
  \bibinfo{pages}{041101}.

\bibitem[{{Beresnyak} et~al.(2005){Beresnyak}, {Dagkesamanskii}, {Zheleznykh},
  {Kovalenko}, and {Oreshko}}]{beresnyak2005}
\bibinfo{author}{A.~R. {Beresnyak}}, \bibinfo{author}{R.~D. {Dagkesamanskii}},
  \bibinfo{author}{I.~M. {Zheleznykh}}, \bibinfo{author}{A.~V. {Kovalenko}},
  \bibinfo{author}{V.~V. {Oreshko}}, \bibinfo{title}{{Limits on the Flux of
  Ultrahigh-Energy Neutrinos from Radio Astronomical Observations}},
  \bibinfo{journal}{\arep{}} \bibinfo{volume}{49} (\bibinfo{year}{2005})
  \bibinfo{pages}{127--133}.

\bibitem[{{James} et~al.(2010){James}, {Ekers}, {Alvarez-Mu{\~n}iz}, {Bray},
  {McFadden}, {Phillips}, {Protheroe}, and {Roberts}}]{james2010}
\bibinfo{author}{C.~W. {James}}, \bibinfo{author}{R.~D. {Ekers}},
  \bibinfo{author}{J.~{Alvarez-Mu{\~n}iz}}, \bibinfo{author}{J.~D. {Bray}},
  \bibinfo{author}{R.~A. {McFadden}}, \bibinfo{author}{C.~J. {Phillips}},
  \bibinfo{author}{R.~J. {Protheroe}}, \bibinfo{author}{P.~{Roberts}},
  \bibinfo{title}{{LUNASKA experiments using the Australia Telescope Compact
  Array to search for ultrahigh energy neutrinos and develop technology for the
  lunar Cherenkov technique}}, \bibinfo{journal}{\prd{}}
  \bibinfo{volume}{81}~(\bibinfo{number}{4}) (\bibinfo{year}{2010})
  \bibinfo{pages}{042003}.

\bibitem[{{Spencer} et~al.(2010){Spencer}, {Macfarlane}, {Mills}, and
  {Piccirillo}}]{spencer2010}
\bibinfo{author}{R.~E. {Spencer}}, \bibinfo{author}{A.~{Macfarlane}},
  \bibinfo{author}{O.~{Mills}}, \bibinfo{author}{L.~{Piccirillo}},
  \bibinfo{title}{{La Luna: Lovell Attempts LUnar Neutrino Acquisition}}, in:
  \bibinfo{booktitle}{Proc.\ EVN Symp.\ 2010}, \procsci{},
  \bibinfo{pages}{097}, \bibinfo{year}{2010}.

\bibitem[{{Buitink} et~al.(2010){Buitink}, {Scholten}, {Bacelar}, {Braun}, {de
  Bruyn}, {Falcke}, {Singh}, {Stappers}, {Strom}, and {Yahyaoui}}]{buitink2010}
\bibinfo{author}{S.~{Buitink}}, \bibinfo{author}{O.~{Scholten}},
  \bibinfo{author}{J.~{Bacelar}}, \bibinfo{author}{R.~{Braun}},
  \bibinfo{author}{A.~G. {de Bruyn}}, \bibinfo{author}{H.~{Falcke}},
  \bibinfo{author}{K.~{Singh}}, \bibinfo{author}{B.~{Stappers}},
  \bibinfo{author}{R.~G. {Strom}}, \bibinfo{author}{R.~A. {Yahyaoui}},
  \bibinfo{title}{{Constraints on the flux of ultra-high energy neutrinos from
  Westerbork Synthesis Radio Telescope observations}},
  \bibinfo{journal}{\aap{}} \bibinfo{volume}{521} (\bibinfo{year}{2010})
  \bibinfo{pages}{A47}.

\bibitem[{{Jaeger} et~al.(2010){Jaeger}, {Mutel}, and {Gayley}}]{jaeger2010}
\bibinfo{author}{T.~R. {Jaeger}}, \bibinfo{author}{R.~L. {Mutel}},
  \bibinfo{author}{K.~G. {Gayley}}, \bibinfo{title}{{Project RESUN, a Radio
  EVLA Search for UHE Neutrinos}}, \bibinfo{journal}{\app{}}
  \bibinfo{volume}{34} (\bibinfo{year}{2010}) \bibinfo{pages}{293--303}.

\bibitem[{{Carilli} and {Rawlings}(2004)}]{carilli2004}
\bibinfo{author}{C.~L. {Carilli}}, \bibinfo{author}{S.~{Rawlings}},
  \bibinfo{title}{{Motivation, key science projects, standards and
  assumptions}}, \bibinfo{journal}{\nar{}} \bibinfo{volume}{48}
  (\bibinfo{year}{2004}) \bibinfo{pages}{979--984}.

\bibitem[{{Greisen}(1966)}]{greisen1966}
\bibinfo{author}{K.~{Greisen}}, \bibinfo{title}{{End to the Cosmic-Ray
  Spectrum?}}, \bibinfo{journal}{\prl{}} \bibinfo{volume}{16}
  (\bibinfo{year}{1966}) \bibinfo{pages}{748--750}.

\bibitem[{{Zatsepin} and {Kuzmin}(1966)}]{zatsepin1966}
\bibinfo{author}{G.~T. {Zatsepin}}, \bibinfo{author}{V.~A. {Kuzmin}},
  \bibinfo{title}{{Upper Limit of the Spectrum of Cosmic Rays}},
  \bibinfo{journal}{\spjetpl{}} \bibinfo{volume}{4} (\bibinfo{year}{1966})
  \bibinfo{pages}{78}.

\bibitem[{{Ekers} et~al.(2009){Ekers}, {James}, {Protheroe}, and
  {McFadden}}]{ekers2009}
\bibinfo{author}{R.~D. {Ekers}}, \bibinfo{author}{C.~W. {James}},
  \bibinfo{author}{R.~J. {Protheroe}}, \bibinfo{author}{R.~A. {McFadden}},
  \bibinfo{title}{{Lunar radio Cherenkov observations of UHE neutrinos}},
  \bibinfo{journal}{\nima{}} \bibinfo{volume}{604} (\bibinfo{year}{2009})
  \bibinfo{pages}{106}.

\bibitem[{{Bray} et~al.(2014){Bray}, {Ekers}, {Roberts}, {Reynolds}, {James},
  {Phillips}, {Protheroe}, {McFadden}, and {Aartsen}}]{parkes_theory}
\bibinfo{author}{J.~D. {Bray}}, \bibinfo{author}{R.~D. {Ekers}},
  \bibinfo{author}{P.~{Roberts}}, \bibinfo{author}{J.~E. {Reynolds}},
  \bibinfo{author}{C.~W. {James}}, \bibinfo{author}{C.~J. {Phillips}},
  \bibinfo{author}{R.~J. {Protheroe}}, \bibinfo{author}{R.~A. {McFadden}},
  \bibinfo{author}{M.~G. {Aartsen}}, \bibinfo{title}{{A limit on the
  ultra-high-energy neutrino flux from lunar observations with the Parkes radio
  telescope}}, \bibinfo{note}{submitted}, \bibinfo{year}{2014}.

\bibitem[{{Staveley-Smith} et~al.(1996){Staveley-Smith}, {Wilson}, {Bird},
  {Disney}, {Ekers}, {Freeman}, {Haynes}, {Sinclair}, {Vaile}, {Webster}, and
  {Wright}}]{staveley-smith1996}
\bibinfo{author}{L.~{Staveley-Smith}}, \bibinfo{author}{W.~E. {Wilson}},
  \bibinfo{author}{T.~S. {Bird}}, \bibinfo{author}{M.~J. {Disney}},
  \bibinfo{author}{R.~D. {Ekers}}, \bibinfo{author}{K.~C. {Freeman}},
  \bibinfo{author}{R.~F. {Haynes}}, \bibinfo{author}{M.~W. {Sinclair}},
  \bibinfo{author}{R.~A. {Vaile}}, \bibinfo{author}{R.~L. {Webster}},
  \bibinfo{author}{A.~E. {Wright}}, \bibinfo{title}{{The Parkes 21 cm multibeam
  receiver}}, \bibinfo{journal}{\pasa{}} \bibinfo{volume}{13}
  (\bibinfo{year}{1996}) \bibinfo{pages}{243--248}.

\bibitem[{{Bray} et~al.(2013{\natexlab{a}}){Bray}, {Ekers}, and
  {Roberts}}]{bray2012}
\bibinfo{author}{J.~D. {Bray}}, \bibinfo{author}{R.~D. {Ekers}},
  \bibinfo{author}{P.~{Roberts}}, \bibinfo{title}{{Noise statistics in a fast
  digital radio receiver: the Bedlam backend for the Parkes radio telescope}},
  \bibinfo{journal}{\expa{}} \bibinfo{volume}{36}~(\bibinfo{number}{1--2})
  (\bibinfo{year}{2013}{\natexlab{a}}) \bibinfo{pages}{155--174}.

\bibitem[{{James} and {Protheroe}(2009{\natexlab{b}})}]{james2009f}
\bibinfo{author}{C.~W. {James}}, \bibinfo{author}{R.~J. {Protheroe}},
  \bibinfo{title}{{The directional dependence of apertures, limits and
  sensitivity of the lunar Cherenkov technique to a UHE neutrino flux}},
  \bibinfo{journal}{\app{}} \bibinfo{volume}{31}
  (\bibinfo{year}{2009}{\natexlab{b}}) \bibinfo{pages}{392--398}.

\bibitem[{{Pierre Auger Collaboration}(2010)}]{abreu2010}
\bibinfo{author}{{Pierre Auger Collaboration}}, \bibinfo{title}{{Update on the
  correlation of the highest energy cosmic rays with nearby extragalactic
  matter}}, \bibinfo{journal}{\app{}} \bibinfo{volume}{34}
  (\bibinfo{year}{2010}) \bibinfo{pages}{314--326}.

\bibitem[{Savitzky and Golay(1964)}]{savitzky1964}
\bibinfo{author}{A.~Savitzky}, \bibinfo{author}{M.~J.~E. Golay},
  \bibinfo{title}{Smoothing and Differentiation of Data by Simplified Least
  Squares Procedures.}, \bibinfo{journal}{\anchem{}}
  \bibinfo{volume}{36}~(\bibinfo{number}{8}) (\bibinfo{year}{1964})
  \bibinfo{pages}{1627--1639}.

\bibitem[{{Moffat}(1972)}]{moffat1972}
\bibinfo{author}{P.~H. {Moffat}}, \bibinfo{title}{{Aperture synthesis
  polarimetry of the Moon at 21~cm}}, \bibinfo{journal}{\mnras{}}
  \bibinfo{volume}{160} (\bibinfo{year}{1972}) \bibinfo{pages}{139--154}.

\bibitem[{{Troitskii} and {Tikhonova}(1970)}]{troitskii1970}
\bibinfo{author}{V.~S. {Troitskii}}, \bibinfo{author}{T.~V. {Tikhonova}},
  \bibinfo{title}{{Thermal radiation from the moon and the physical properties
  of the upper lunar layer}}, \bibinfo{journal}{\rqe{}} \bibinfo{volume}{13}
  (\bibinfo{year}{1970}) \bibinfo{pages}{981--1010}.

\bibitem[{{Roberts}(2008)}]{roberts2008}
\bibinfo{author}{P.~{Roberts}}, \bibinfo{title}{Ionospheric dispersion
  compensation using a novel microwave de-dispersion filter}, in:
  \bibinfo{booktitle}{Proc.\ WARS 2008}, \bibinfo{note}{{arXiv:1305.0333}},
  \bibinfo{year}{2008}.

\bibitem[{Hern\'{a}ndez-Pajares et~al.(2009)Hern\'{a}ndez-Pajares, Juan, Sanz,
  Orus, Garcia-Rigo, Feltens, Komjathy, Schaer, and
  Krankowski}]{hernandez-pajares2009}
\bibinfo{author}{M.~Hern\'{a}ndez-Pajares}, \bibinfo{author}{J.~Juan},
  \bibinfo{author}{J.~Sanz}, \bibinfo{author}{R.~Orus},
  \bibinfo{author}{A.~Garcia-Rigo}, \bibinfo{author}{J.~Feltens},
  \bibinfo{author}{A.~Komjathy}, \bibinfo{author}{S.~Schaer},
  \bibinfo{author}{A.~Krankowski}, \bibinfo{title}{The IGS VTEC maps: a
  reliable source of ionospheric information since 1998},
  \bibinfo{journal}{\jgeod{}} \bibinfo{volume}{83} (\bibinfo{year}{2009})
  \bibinfo{pages}{263--275}.

\bibitem[{{Noll}(2010)}]{noll2010}
\bibinfo{author}{C.~{Noll}}, \bibinfo{title}{The crustal dynamics data
  information system: A resource to support scientific analysis using space
  geodesy}, \bibinfo{journal}{\aspacer{}}
  \bibinfo{volume}{45}~(\bibinfo{number}{12}) (\bibinfo{year}{2010})
  \bibinfo{pages}{1421--1440}.

\bibitem[{{Schaer} et~al.(1998){Schaer}, {Gurtner}, and {Feltens}}]{schaer1998}
\bibinfo{author}{S.~{Schaer}}, \bibinfo{author}{W.~{Gurtner}},
  \bibinfo{author}{J.~{Feltens}}, \bibinfo{title}{{IONEX: The IONosphere Map
  EXchange Format Version 1}}, in: \bibinfo{booktitle}{{Proc.\ IGS Analysis
  Center Workshop 1998}}, \bibinfo{pages}{233--247}, \bibinfo{year}{1998}.

\bibitem[{{Gallagher} et~al.(1988){Gallagher}, {Craven}, and
  {Comfort}}]{gallagher1988}
\bibinfo{author}{D.~L. {Gallagher}}, \bibinfo{author}{P.~D. {Craven}},
  \bibinfo{author}{R.~H. {Comfort}}, \bibinfo{title}{{An empirical model of the
  earth's plasmasphere}}, \bibinfo{journal}{\aspacer{}} \bibinfo{volume}{8}
  (\bibinfo{year}{1988}) \bibinfo{pages}{15--21}.

\bibitem[{Byrne(2005)}]{byrne2005}
\bibinfo{author}{C.~L. Byrne}, \bibinfo{title}{Signal processing: a
  mathematical approach}, \bibinfo{publisher}{A.\ K.\ Peters}, ISBN
  \bibinfo{isbn}{1-56881-242-6}, \bibinfo{year}{2005}.

\bibitem[{{Alvarez-Mu{\~n}iz} et~al.(2006){Alvarez-Mu{\~n}iz}, {Marqu{\'e}s},
  {V{\'a}zquez}, and {Zas}}]{alvarez-muniz2006}
\bibinfo{author}{J.~{Alvarez-Mu{\~n}iz}}, \bibinfo{author}{E.~{Marqu{\'e}s}},
  \bibinfo{author}{R.~A. {V{\'a}zquez}}, \bibinfo{author}{E.~{Zas}},
  \bibinfo{title}{{Coherent radio pulses from showers in different media: A
  unified parametrization}}, \bibinfo{journal}{\prd{}}
  \bibinfo{volume}{74}~(\bibinfo{number}{2}) (\bibinfo{year}{2006})
  \bibinfo{pages}{023007}.

\bibitem[{{Mio{\v c}inovi{\'c}} et~al.(2006){Mio{\v c}inovi{\'c}}, {Field},
  {Gorham}, {Guillian}, {Milin{\v c}i{\'c}}, {Saltzberg}, {Walz}, and
  {Williams}}]{miocinovic2006}
\bibinfo{author}{P.~{Mio{\v c}inovi{\'c}}}, \bibinfo{author}{R.~C. {Field}},
  \bibinfo{author}{P.~W. {Gorham}}, \bibinfo{author}{E.~{Guillian}},
  \bibinfo{author}{R.~{Milin{\v c}i{\'c}}}, \bibinfo{author}{D.~{Saltzberg}},
  \bibinfo{author}{D.~{Walz}}, \bibinfo{author}{D.~{Williams}},
  \bibinfo{title}{{Time-domain measurement of broadband coherent Cherenkov
  radiation}}, \bibinfo{journal}{\prd{}}
  \bibinfo{volume}{74}~(\bibinfo{number}{4}) (\bibinfo{year}{2006})
  \bibinfo{pages}{043002}.

\bibitem[{{Alvarez-Mu{\~n}iz} et~al.(2010){Alvarez-Mu{\~n}iz}, {Romero-Wolf},
  and {Zas}}]{alvarez-muniz2010}
\bibinfo{author}{J.~{Alvarez-Mu{\~n}iz}}, \bibinfo{author}{A.~{Romero-Wolf}},
  \bibinfo{author}{E.~{Zas}}, \bibinfo{title}{{{\v C}erenkov radio pulses from
  electromagnetic showers in the time domain}}, \bibinfo{journal}{\prd{}}
  \bibinfo{volume}{81}~(\bibinfo{number}{12}) (\bibinfo{year}{2010})
  \bibinfo{pages}{123009}.

\bibitem[{{Alvarez-Mu{\~n}iz} et~al.(2000){Alvarez-Mu{\~n}iz}, {V{\'a}zquez},
  and {Zas}}]{alvarez-muniz2000}
\bibinfo{author}{J.~{Alvarez-Mu{\~n}iz}}, \bibinfo{author}{R.~A.
  {V{\'a}zquez}}, \bibinfo{author}{E.~{Zas}}, \bibinfo{title}{{Calculation
  methods for radio pulses from high energy showers}},
  \bibinfo{journal}{\prd{}} \bibinfo{volume}{62}~(\bibinfo{number}{6})
  (\bibinfo{year}{2000}) \bibinfo{pages}{063001}.

\bibitem[{{Longuet-Higgins}(1957)}]{longuet-higgins1957}
\bibinfo{author}{M.~S. {Longuet-Higgins}}, \bibinfo{title}{{The Statistical
  Analysis of a Random, Moving Surface}}, \bibinfo{journal}{\rsla{}}
  \bibinfo{volume}{249} (\bibinfo{year}{1957}) \bibinfo{pages}{321--387}.

\bibitem[{{Gaisser} and {Hillas}(1977)}]{gaisser1977}
\bibinfo{author}{T.~K. {Gaisser}}, \bibinfo{author}{A.~M. {Hillas}},
  \bibinfo{title}{{Reliability of the method of constant intensity cuts for
  reconstructing the average development of vertical showers}}, in:
  \bibinfo{booktitle}{Proc.\ ICRC 1977}, vol.~\bibinfo{volume}{8},
  \bibinfo{pages}{353--357}, \bibinfo{year}{1977}.

\bibitem[{{Gayley} et~al.(2009){Gayley}, {Mutel}, and {Jaeger}}]{gayley2009}
\bibinfo{author}{K.~G. {Gayley}}, \bibinfo{author}{R.~L. {Mutel}},
  \bibinfo{author}{T.~R. {Jaeger}}, \bibinfo{title}{{Analytic Aperture
  Calculation and Scaling Laws for Radio Detection of Lunar-target
  Ultra-high-energy Neutrinos}}, \bibinfo{journal}{\apj{}}
  \bibinfo{volume}{706} (\bibinfo{year}{2009}) \bibinfo{pages}{1556--1570}.

\bibitem[{{Singh} et~al.(2012){Singh}, {Mevius}, {Scholten}, {Anderson}, {van
  Ardenne}, {Arts}, {Avruch}, {Asgekar}, {Bell}, {Bennema}, and
  et~al.}]{singh2012}
\bibinfo{author}{K.~{Singh}}, \bibinfo{author}{M.~{Mevius}},
  \bibinfo{author}{O.~{Scholten}}, \bibinfo{author}{J.~M. {Anderson}},
  \bibinfo{author}{A.~{van Ardenne}}, \bibinfo{author}{M.~{Arts}},
  \bibinfo{author}{M.~{Avruch}}, \bibinfo{author}{A.~{Asgekar}},
  \bibinfo{author}{M.~{Bell}}, \bibinfo{author}{P.~{Bennema}},
  \bibinfo{author}{et~al.}, \bibinfo{title}{{Optimized trigger for
  ultra-high-energy cosmic-ray and neutrino observations with the low frequency
  radio array}}, \bibinfo{journal}{\nima{}} \bibinfo{volume}{664}
  (\bibinfo{year}{2012}) \bibinfo{pages}{171--185}.

\bibitem[{{Bray} et~al.(2013{\natexlab{b}}){Bray}, {Ekers}, {Protheroe},
  {James}, {Phillips}, {Roberts}, {Brown}, {Reynolds}, {McFadden}, and
  {Aartsen}}]{bray2013}
\bibinfo{author}{J.~D. {Bray}}, \bibinfo{author}{R.~D. {Ekers}},
  \bibinfo{author}{R.~J. {Protheroe}}, \bibinfo{author}{C.~W. {James}},
  \bibinfo{author}{C.~J. {Phillips}}, \bibinfo{author}{P.~{Roberts}},
  \bibinfo{author}{A.~{Brown}}, \bibinfo{author}{J.~E. {Reynolds}},
  \bibinfo{author}{R.~A. {McFadden}}, \bibinfo{author}{M.~{Aartsen}},
  \bibinfo{title}{{LUNASKA neutrino search with the Parkes and ATCA
  telescopes}}, in: \bibinfo{booktitle}{Proc.\ ARENA 2012}, vol.
  \bibinfo{volume}{1535} of \emph{\bibinfo{series}{\aipcs{}}},
  \bibinfo{pages}{21--26}, \bibinfo{year}{2013}{\natexlab{b}}.

\bibitem[{{Schinckel} et~al.(2012){Schinckel}, {Bunton}, {Cornwell}, {Feain},
  and {Hay}}]{schinckel2012}
\bibinfo{author}{A.~E. {Schinckel}}, \bibinfo{author}{J.~D. {Bunton}},
  \bibinfo{author}{T.~J. {Cornwell}}, \bibinfo{author}{I.~{Feain}},
  \bibinfo{author}{S.~G. {Hay}}, \bibinfo{title}{{The Australian SKA
  Pathfinder}}, in: \bibinfo{booktitle}{Ground-based and Airborne Telescopes
  IV}, vol. \bibinfo{volume}{8444} of \emph{\bibinfo{series}{\procspie{}}},
  \bibinfo{pages}{84442A}, \bibinfo{year}{2012}.

\bibitem[{{McFadden} et~al.(2011){McFadden}, {Ekers}, and
  {Bray}}]{mcfadden2011}
\bibinfo{author}{R.~A. {McFadden}}, \bibinfo{author}{R.~D. {Ekers}},
  \bibinfo{author}{J.~D. {Bray}}, \bibinfo{title}{{Ionospheric propagation
  effects for UHE neutrino detection with the lunar Cherenkov technique}}, in:
  \bibinfo{booktitle}{Proc.\ ICRC 2011}, vol.~\bibinfo{volume}{4},
  \bibinfo{pages}{284}, \bibinfo{year}{2011}.

\end{thebibliography}

\end{document}